\documentclass[twocolumn]{aastex63}

\usepackage{hyperref}
\usepackage{amsmath}
\usepackage{mathrsfs}
\usepackage[]{units} 
\usepackage{threeparttable}
\usepackage{multirow}
\usepackage{soul}
\usepackage{ulem}
\usepackage{graphicx}
\usepackage{xspace}

\def\RTsim{{\tt {RTsim}}}
\def\nonRTsim{{\tt nonRTsim}}

\begin{document}

\title{The Effects of Radiative Feedback and Supernova Induced Turbulence on Early Galaxies}
\shorttitle{Radiative Feedback \& Structure Formation}
\author{Richard Sarmento}
\affiliation{Dept. of Computer Science, United States Naval Academy,  121 Blake Road, Annapolis, MD, 21402, USA}
\affiliation{School of Earth and Space Exploration, 
Arizona State University, 
P.O. Box 871404, Tempe, AZ, 85287-1404}

\author{Evan Scannapieco}
\affiliation{School of Earth and Space Exploration, 
Arizona State University, 
P.O. Box 871404, Tempe, AZ, 85287-1404}

\shortauthors{Sarmento and Scannapieco}

\begin{abstract}

The recently launched {\em James Webb Space Telescope} promises unparalleled advances in our understanding of the first stars and galaxies, but realizing this potential requires cosmological simulations that capture the key physical processes that affected these objects. Here we show that radiative transfer and subgrid turbulent mixing are two such processes.  By comparing simulations with and without radiative transfer but with exactly the same physical parameters and subgrid turbulent mixing model,  we show that tracking radiative transfer suppresses the Population III  (Pop III) star formation density by a factor $\approx 4.$  In both simulations,  $\gtrsim 90\%$ of Pop III stars are found in the unresolved pristine regions tracked by our subgrid model, which does a better job at modeling the regions surrounding proto-galaxy cores where metals from supernovae take tens of Myrs to mix thoroughly.   At the same time, radiative transfer suppresses Pop III star formation, via the development of ionized bubbles that slows gas accretion in these regions, and it results in compact high-redshift galaxies that are surrounded by isolated low mass satellites.  Thus turbulent mixing and radiative transfer are both essential processes that must be included to accurately model the morphology, composition, and growth of primordial galaxies. 

\end{abstract}


\section{Introduction}

The search to observe the first stars and galaxies is one of the most active frontiers in astronomy. Using the 
{\em James Webb Space Telescope} ({\em JWST}), researchers will soon be able to measure galaxies out to redshift 10 and beyond, probing the era in which the first generation of stars was formed \citep{Gardner_2006}. As such, it is important to have physical models that predict the density of early galaxies well beyond the redshift and luminosity limits of current surveys \citep[e.g.][]{Finkelstein_2016, Ishigaki2018a, Bouwens2019}. 

The first stars were formed of primordial gas made up of hydrogen, helium, and trace amounts of lithium, and are believed to have been much more massive than the sun \citep[e.g.][]{abel2002formation,Bromm2002,2007ApJ...654...66O,Susa2014}. While the very first stars were composed solely of primordial gas, it is also believed that stars with traces of heavier elements less than a critical metallicity   $Z_{\rm crit}$, also share their characteristics \citep{Schneider2011}. These Population III (Pop III) stars have a strong impact on the luminosity of early galaxies \citep{2017MNRAS.467L..51Y} due to their lower opacity, higher surface temperatures, and enhanced UV spectra \citep{Raiter2010}, and may even be able to be observed individually through the use of strong lensing \citep[e.g.][]{Windhorst2018,Welch:2022tp}.

However, the physical processes governing the formation and characteristics of Pop III stars are poorly understood \citep{Ishiyama_2016}. 
The lack of confirmed, direct observations of Pop III stars has led to an abundance of numerical and theoretical studies \citep[e.g.][]{Mackey2003,2003ApJ...589...35S,2007ApJ...654...66O,Ahn2007,2012ApJ...745...50W,2014MNRAS.440.2498P,Xu2016}, each of which focuses on a particular subset of the relevant physics. Important among these are the models for radiative and supernova (SN) feedback, both of which heat the gas, helping it resist collapse.  However, SN feedback also pollutes the gas with heavy elements that help gas cool more efficiently \citep{Hirano_2013} and begins the transition to lower-mass Population II (Pop II) star formation.  

The pollution of pristine gas depends on at least two parameters: the critical metallicity and the rate at which metals ejected by SN are distributed throughout early star-forming halos \citep{2013ApJ...775..111P}. The first parameter is poorly understood but is believed to lie between $10^{-6} {Z_{\odot}}$ and  $10^{-3} Z_{\odot}$ \citep{Schneider2011,2013ApJ...766..103D}. This is a range of values that is low enough such that the uncertainty in $Z_{\rm crit}$ has a very weak impact on predictions of the evolution of the Pop III stars \citep{Sarmento2019}. 

Instead, the transition from Pop III to Pop II star formation is governed by another critical component: the transport of SN ejecta into the primordial gas. This process determines the degree to which a specific parcel of gas is polluted with heavy elements and hence whether it will subsequently form metal-free Pop III or metal-enriched Pop II stars \citep{Sarmento2016}.  Since metallicity relates to cooling, the level of pollution also dictates the star formation timescale.  Thus the combined effects of supernovae and turbulent mixing led to a spatially-inhomogeneous Pop III/Pop II transition, the details of which are key to understanding  the structure and evolution  of the earliest galaxies \citep{2003ApJ...589...35S,2012ApJ...745...50W,2013ApJ...773..108C,2013MNRAS.428.1857J,2013ApJ...775..111P,2014MNRAS.440.2498P}.

Our previous work has emphasized the importance of turbulent 
mixing in this transition, developing a new approach that allows us to track the effects of subgrid mixing in each resolution element.   By using a self-convolution model developed in  \cite{2012JFM...700..459P,2013ApJ...775..111P}
 to estimate the rate at which turbulence mixes pollutants,
we were able to show that thorough mixing can take several eddy turnover
times.  This means that modeling turbulent mixing is at least as important as modeling large-scale inhomogeneities when determining the Pop III star formation history. In fact, we found an increase of $\approx 3$ times the Pop III star-formation rate density as compared to similar cosmological simulations that did not account for subgrid mixing \citep{Sarmento2016} and were able to account for this effect to make several detailed predictions as to the evolution of the first stars and galaxies \citep{Sarmento2018,Sarmento2019}.

However, this picture is further complicated by the presence of radiative feedback. Pop III stars and massive Pop II stars produce copious amounts of ultraviolet radiation in both the H-ionizing and Lyman-Werner (LW) bands, which has a significant impact on the rate of  star formation.  LW radiation dissociates molecular hydrogen, which is the primary coolant of primordial gas and the predominant cooling channel for Pop III star formation \citep{Ahn2009,Safranek-Shrader2018}.  Hydrogen reionizing photons, on the other hand, provide an additional heating source that inhibits the formation of stars in low-mass galaxies \citep{Dawoodbhoy2018,Katz2020}.

Dedicated radiative-transfer simulations have studied reionization in detail, tracking the propagation of ionization fronts outward from early galaxies and continuing until the HII regions bounded by these fronts overlap, fully ionizing the universe.  These were initially carried out by post-processing cosmological density fields from simulations of large-scale structure formation \citep[e.g.][]{Abel1999,Gnedin2000,Nakamoto2001,Ciardi2001,Ciard2006,Sokasian2004,Zahn2011}, but
later these advanced to the point where gas dynamics and radiative-transfer could be simulated
self-consistently
\citep[e.g.][]{Iliev2007,Gnedin2014,Ocvirk2015,Pawlik2017,Ocvirk2020}.
Together these simulations described an extended process by which HII regions first formed around the highly-clustered densest sources, grew in size and number and became more-aspherical with time, and finally percolated to ionize the remaining islands of neutral material \citep[e.g.][]{Furlanetto2006,Zahn2007}.

Several authors have performed simulations to model the feedback effects of  both SN and ionizing radiation on the Pop III/II SFR. Many of these models employed a simplified heating or global radiation model \citep[e.g.][]{2007MNRAS.382..945T, Johnson2008, Greif2008,Jaacks2018} while others employed a self-consistent locally generated radiation field \citep[e.g.][]{2012ApJ...745...50W,Pawlik2013,2013MNRAS.428.1857J,Wells2021}. However, no one has yet combined self-consistent radiative transfer and SN feedback with a subgrid turbulent mixing model for SN ejecta.

This work describes the impact of modeling radiative transfer in the context of the very high-redshift universe. To that end, we describe the differences between two pilot simulations. The \RTsim{} tracks a self-consistently generated radiation field coupled with SN feedback and turbulent mixing.  The \nonRTsim{} replaces radiative transfer with a homogeneous UV background \citep{1996ApJ...461...20H} switched on at $z=9$, while still retaining SN feedback and turbulent mixing. Together these simulations allow us to tease out the effects of radiative feedback on the formation of the first galaxies.

The work is structured as follows.  In Section 2, we describe our methods including the spectral energy distribution (SED) models used to model radiation from our stars. In Section 3, we compare the \nonRTsim{} and \RTsim{} simulations, focusing on the impact of radiative transfer on the overall evolution and the spatial distribution of Pop III stars. Conclusions are discussed in Section 4. 

\section{Methods}

Our simulations use a modified version of \textsc{Ramses-RT} \citep{Rosdahl2013, Rosdahl2015a}, a cosmological adaptive mesh refinement (AMR) simulation with coupled radiation hydrodynamics (RHD), to which we added our specialized model of turbulent mixing.  Here we describe the key features and parameters of the simulations.

\subsection{Radiative Transfer and Turbulent Mixing}

\textsc{Ramses-RT} is an extension of \textsc{Ramses} \citep{Teyssier2001} that models the interactions between dark matter, stellar populations, and baryons via gravity and hydrodynamics. \textsc{Ramses-RT} adds stellar radiation and radiative transfer as well as non-equilibrium radiative heating and cooling. The simulation advects photons between cells using a first-order moment method with full local M1 closure for the Eddington tensor \citep{LEVERMORE1984149}.
To keep the radiative transfer computations manageable, we group photons into four energy ranges: the H$_2$ dissociating LW band, H ionizing radiation, and 2 levels for He ionizing radiation, which correspond to
\begin{align*}
 11.20\,{\rm eV} &\leq \epsilon_{\rm LW} < 13.60\,{\rm eV}, \\
 13.60\,{\rm eV} &\leq \epsilon_{\rm HII} < 24.59\,{\rm eV},\\
 24.59\,{\rm eV} &\leq \epsilon_{\rm HeII} < 54.42\,{\rm eV}, \,\, {\rm and} \\
 54.42\,{\rm eV} &\leq \epsilon_{\rm HeIII}.
\end{align*}  

\textsc{Ramses-RT} models the thermochemistry, photon absorption, and emission \citep{Rosdahl2013,Rosdahl_2015,Nickerson2018}. The ionization states are modeled self-consistently with the temperature and radiation field in terms of photon density and flux for each of the photon groups, and the ionization fractions $x_{\rm HII}$, $x_{\rm HeII}$, $x_{\rm HeIII}$ are computed, stored, and advected between cells. The photon densities and fluxes, $N_i$ and $F_i$ for each photon group, are computed using moment-based radiative-transfer that essentially treats the photons as a fluid. Source photon abundances are generated for star particles (SPs) using an externally-specified spectral energy density (SED) model. We use {\it Starburst99} \citep{2011ascl.soft04003L} and \cite{Raiter2010} to model stellar radiation from populations with metallicities $0 \le Z \le Z_{\odot}$ and ages between 10 kyr and a Gyr. Photon energies are also discretized across our 4 bins, resulting in 4 average photon energies.  The simulation employs a reduced speed of light approx \citep{Ocvirk_2019} which greatly increases the allowable time step, and we adopt 0.01$c$ for this value.

The hydrodynamic flux between cells is computed using a Harten--Lax--van Leer contact (HLLC) Riemann solver \citep{Toro:1994we}. It is used to advect the typical cell-centered gas variables, the RT ionization states, as well as the hydro scalars added by \cite{Sarmento2016} that track the turbulent velocity, the pristine gas mass fraction, and the metals generated by Population III (Pop III) supernova (SNe). 

We use the self convolution model developed by \cite{2013ApJ...775..111P} to follow subgrid mixing of SN ejecta into the pristine gas. The fraction of pristine gas in each cell is modeled as a scalar value $P$ that evolves as
\begin{align}\label{eq:selfConv}
\frac{d P}{d t} = - \frac{n}{\tau_{\rm con}}  P\left(1-P ^{1/n}\right), 
\end{align}
where $\tau_{\rm con}$ is the convolution timescale that is inversely proportional to the turbulent stretching rate \citep{2010ApJ...721.1765P, 2012JFM...700..459P, 2013ApJ...775..111P} and $n$ is a measure of the locality of mixing. The scalar $P$ is tracked for each cell and computed at each time step when $P>0$.

By modeling the unmixed fraction of gas in each simulation cell, we can estimate the fraction of Pop III vs. Pop II stars formed in these actively mixing regions. This leads to an enhanced Pop III star formation rate density  as compared to the rate derived in simulations that do not track $P,$ and assume that Pop III stars form in cells composed solely of primordial gas. We adopt the name ``Classical Pop III" for stars formed in these regions, to distinguish them from the overall Pop III stellar distribution, which also includes stars formed in polluted cells with nonzero pristine fractions.
\subsection{Star Formation}
Stars are modeled as collisionless particles and are evolved using a particle-mesh solver with cloud-in-cell interpolation \citep{Guillet_2011}. We assume an ideal gas with a ratio of specific heats $\gamma$ = 5/3. Star particles (SPs) are created in regions of gas according to a  \cite{Schmidt59} law with 
\begin{equation}\label{eqn:sf}
\dot{\rho_{\star}} =  \epsilon_{\star} \frac{\rho_{\rm gas}}{t_{\rm ff}} \theta(\rho_{\rm gas}- \rho_{\rm th}),
\end{equation}
where  $\epsilon_{\star}=0.10$ is the star formation efficiency, $t_{\rm ff} = \sqrt{3 \pi /(32 G \rho)}$ is the gas free-fall time, $\rho_{\rm gas}$ is the local gas density, and
 the Heaviside step function, $\theta$, guarantees star formation occurs only when the gas density $\rho_{\rm gas}$ also exceeds a threshold value $\rho_{\rm th}.$ Here we set $\rho_{\rm th}$ to the larger of  $0.05\, H\, {\rm cm}^{-3}$ and $200\, \overline{\rho}$ where $\overline{\rho}$ is the mean gas density of the simulation. The former criteria is derived from the Jean's condition such that $4 \Delta x_{best} \ge \lambda_j$, where $\lambda_j$ is the Jean's length and $\Delta x_{best}$ is the highest resolution cell size. The latter criteria ensures star formation only occurs in collapsed objects \citep{Rasera2006, Trebitsch2017}. Our $\rho_{\rm th}$  and $\epsilon_{\star}$ result in a star formation rate density (SFRD) in reasonable agreement with observations \citep{Finkelstein_2016, Madau_2014}, see Fig.~\ref{fig:sfrd}, while ensuring we do not form stars in high-density regions of the cosmological flow.

Each SP represents an initial mass function (IMF) of stars. The SP mass is set by the star-forming density threshold and our resolution, resulting in $m_{\star} = (0.05\, H {\rm cm}^{-3}) \Delta x^{3}_{\rm ave} \approx 2.6 \times 10^{3}\, M_{\odot}$. The final mass of each SP is drawn from a Poisson process such that it is a multiple of $m_{\star}$.

\subsection{Feedback}\label{sec:feedback}
For Pop II stars ($Z > Z_{\rm crit}$), we assume a Salpeter IMF such that 10\% of each SP's mass represents stars more massive than 8 $M_\odot$ and go supernova in 10 Myr \citep[e.g.][]{Raskin_2008, Somerville2008}. For Pop III SPs ($Z \le Z_{\rm crit}$), we assume a log-normal, top-heavy IMF such that 99\%, by mass, explode within 10 Myr \citep{1973MNRAS.161..133L, tumlinson2006chemical, Raiter2010} of the SP's formation. These fractions also represent the fractions of SP mass recycled into the ISM for Pop II, $\eta_{\rm sn}=0.10$, and Pop III SPs, $\eta_{\rm snIII}=0.99$, respectively.

The impact of these SNe are parameterized by the mass fraction of ejecta, described above, and the kinetic energy per unit mass of the explosion, $E_{\rm SN}$. We use $E_{\rm SN}=10^{51}$ ergs/10 $M_{\odot}$ for all stars formed throughout the simulation. The fraction of new metals in SN ejecta is 0.15 even though metal yields and energy from Pop III SNe are likely to have been higher \citep{2003ApJ...589...35S,2005ApJ...624L...1S}.

As discussed above, \textsc{Ramses-RT} tracks the ionization states of H$_2$, H, and He, along with the radiation field to compute heating and cooling. The model self-consistently follows collisional and photo-induced ionization, photo-dissociation, and recombination, but does not include cosmic-ray heating. Lastly, we use the on-the-spot approximation \citep{Rosdahl2013}, such that  recombination photons are reabsorbed in the same cell in which they are emitted.

The \nonRTsim{} uses standard \textsc{Ramses} heating and cooling \citep{Teyssier2001} with the addition of a simple molecular cooling model described in \cite{Sarmento2016}. The non-RT H$_2$ cooling is only effective before the first stars form since they generate enough LW photons to dissociate all of the H$_2$ within the \nonRTsim{} box \citep[e.g.][]{2006MNRAS.373..128G}. The \nonRTsim{} model is initialized with a constant H$_2$ fraction, $f_{\rm H_2} = 10^{-6}$, \citep{Reed2005} where-as the \RTsim{} initializes the H$_2$ fraction from the simulation's starting temperature and density.

The photon escape fraction is set to $f_{\star, {\rm esc}} = 0.5$. This parameter scales the fraction of stellar photons that escape the local star-forming cell. Note that $f_{\star, {\rm esc}}$ is distinct from the \textit{galactic} escape fraction that is highly uncertain in the early universe \citep{Tanaka2020}. The $f_{\star, {\rm esc}}$ parameter in the simulation is used to scale SED radiation \citep{Rosdahl2018}. A value $f_{\star, {\rm esc}} < 1$ is typically interpreted as accounting for unresolved over-densities or underestimated recombination rates that trap radiation locally. A value of $f_{esc} > 1$ is typically used to boost stellar luminosities and compensate for chimneys and unresolved turbulence in the inter-stellar/galactic medium. As such, $f_{\star, {\rm esc}}$ is largely a free parameter used to tune RT simulations. We will explore its effects on the overall SFRD in future work.

Finally, we did not include black holes (BH) in our simulation since BH feedback is not likely to be significant for our very early galaxies \citep{Scannapieco_2004,Somerville2008}.  For the \nonRTsim{}, we turn on the UV background at $z=9$, the point at which the \RTsim{} approaches 50\% reionization. 

\subsection{Simulation Parameters}

We adopt a set of cosmological parameters from \cite{Komatsu_2011} with $\Omega_{\rm m} = 0.267$, $\Omega_{\Lambda} = 0.733$, $\Omega_{\rm b} = 0.0449$, $h = 0.71$, $\sigma_8 = 0.801$, and $n = 0.96,$  where $\Omega_{\rm m}$, $\Omega_{\Lambda}$, and $\Omega_{\rm b}$ are the total matter, vacuum, and baryonic densities, in units of the critical density,  $h$ is the Hubble constant in units of 100 km/s; $\sigma_8$ is the variance of linear fluctuations on the 8 $h^{-1}$ Mpc scale; and $n$ is the ``tilt" of the primordial power spectrum \citep{Larson_2011}. 

We evolve two 3 $h^{-1}$ comoving Mpc (cMpc) on-a-side simulations to $z=6$. This is approximately the volume of the local Milky Way (MW) Group. The \RTsim{} includes feedback from stellar photons and radiative transfer while the \nonRTsim{} does not. Otherwise, the two simulations share all common parameters described below.  Note that the small size of our simulation volumes means that they will not capture the large-scale fluctuations due to the patchy structure of reionization, which occurs on scales of 10 Mpc or larger \citep{Furlanetto2004} and requires simulations boxes at least 100 Mpc across to be adequately reproduced \citep{Iliev2014}.  On the other hand, our simulations are well suited to draw conclusions about the {\em small-scale}  features of radiative feedback on the very low mass galaxies within which the first stars were formed.

We set the initial refinement level to $\ell_{\rm min} = 9,$ corresponding to a coarse (initial) grid resolution $\rm \Delta x_{\rm max} = 5.86\; h^{-1}$ comoving kpc (ckpc) -- a compromise that provides reasonable resolution of the intergalactic medium (IGM) without creating an excessive computational load. We adopt a quasi-Lagrangian approach to refinement such that cells are refined as they become approximately 8x over-dense. This strategy attempts to keep the amount of mass in each cell roughly constant as the simulation progresses. 

Allowing for up to 6 additional levels of refinement results in a best average resolution of 91.6 $\rm h^{-1}$ comoving pc (cpc) in refined regions. The initial grid scale and simulation size sets the dark matter (DM) particle mass. For this simulation $M_{\rm DM} = 1.75 \times 10^{4}\, M_{\odot}$ resulting in approximately 300 DM particles needed to resolve a minihalo of $\approx 5.25\times 10^6 M_\odot$ -- a mass near the atomic cooling limit \citep{Wise2014}.

Initial conditions were identical for both simulations and were generated using Multi-Scale Initial Conditions \citep[MUSIC, ][]{2013ascl.soft11011H}. The initial gas metallicity was $Z = 0$ and we define $Z_{\rm crit} = 10^{-5} Z_\odot$, the boundary between Pop III and Population II (Pop II) star formation. The nonlinear length scale at the end of the simulation, $z= 6$, was 39 h$^{-1}$ ckpc , corresponding to a mass of $1.8\times 10^{7}$ h$^{-1}$ $M_\odot$. 

\section{Results}

\begin{figure}[t]
\begin{center}
\includegraphics[width=1\columnwidth]{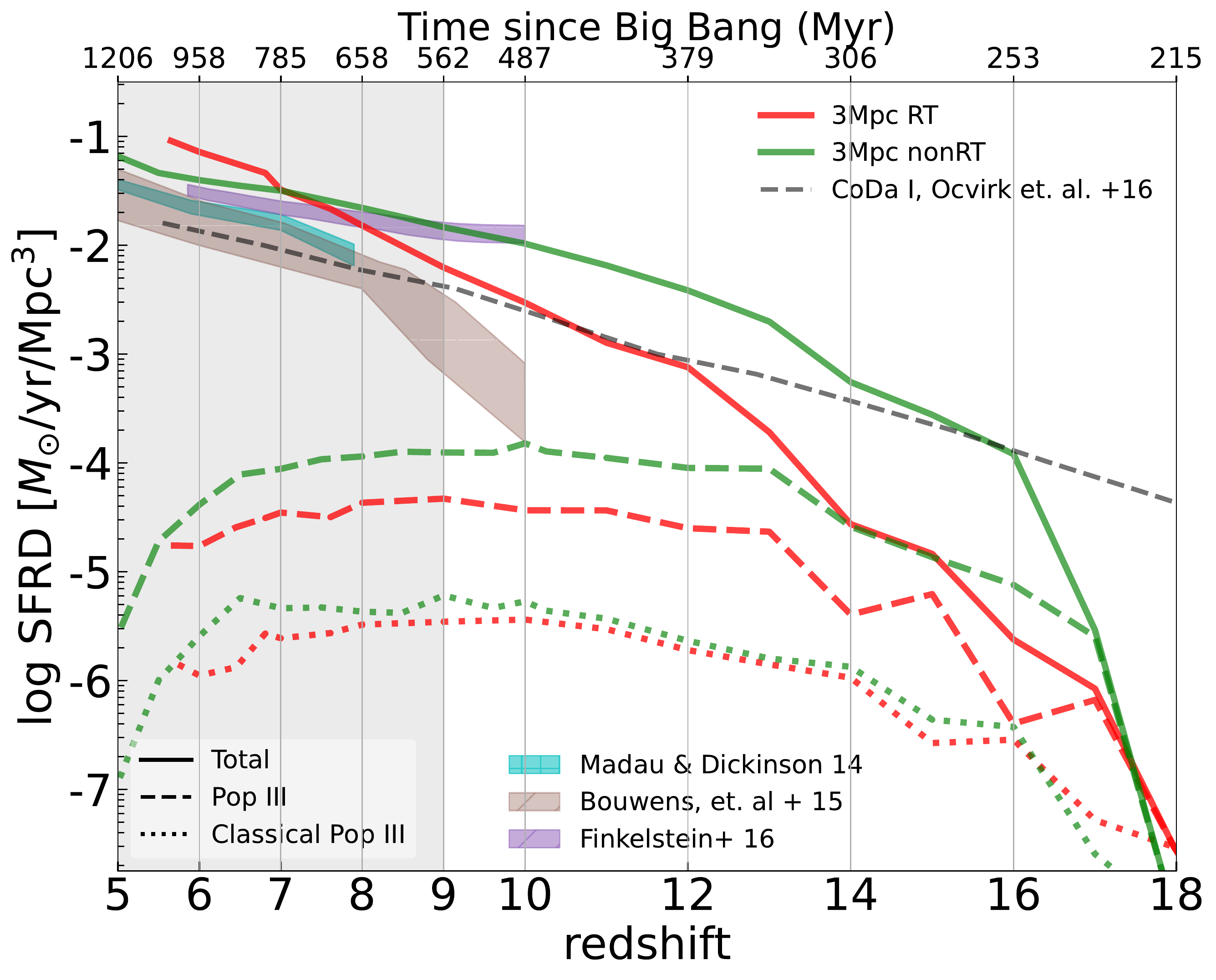}
\caption{The star formation rate density (SFRD) for the \nonRTsim{} (green) and the \RTsim{} (red). The gray region indicates the presence of a \cite{1996ApJ...461...20H} UV background in the \nonRTsim{}. The \nonRTsim{} quickly out-paces the \RTsim{} SFRD at high redshift. However, both simulations generate approximately the same mass of Classical Pop III stars (dotted line). These Pop III stars are created only in simulation cells made up completely of primordial gas. When including stars formed in unresolved primordial regions the \nonRTsim{} generates an average of 4 times more Pop III stars (dashed lines) than the \RTsim{}.}
\label{fig:sfrd}
\end{center}
\end{figure}

\subsection{Overall Evolution}

While our volume is too small to characterize the high-redshift luminosity function, it is well suited to draw conclusions about the impact of modeling RT and turbulent mixing on some of the properties of early galaxies.  As shown in Fig.~\ref{fig:sfrd}, star formation in the \RTsim{} begins slightly earlier than in the \nonRTsim{}.  This is due to the more efficient non-equilibrium cooling used in the \RTsim{} \citep{Rosdahl2013} coupled with different starting H$_2$ fractions for the two simulations, as discussed above. However, although the initial conditions are slightly more favorable for early star formation in the \RTsim{}, by $z=17$ it generates stars at a much slower rate than in the \nonRTsim{}.  This is because
 stellar radiation adds energy to star-forming regions in the \RTsim{}, heating the gas and greatly impeding its ability to collapse and form stars \citep{2013MNRAS.428.1857J,Pawlik2013,Hopkins2020}. Between  $10 < z < 18$ the \nonRTsim{} generates $\approx 9\times$ more mass in SPs than the \RTsim{}.

\begin{figure}[t]
\begin{center}
\includegraphics[width=1.025\columnwidth]{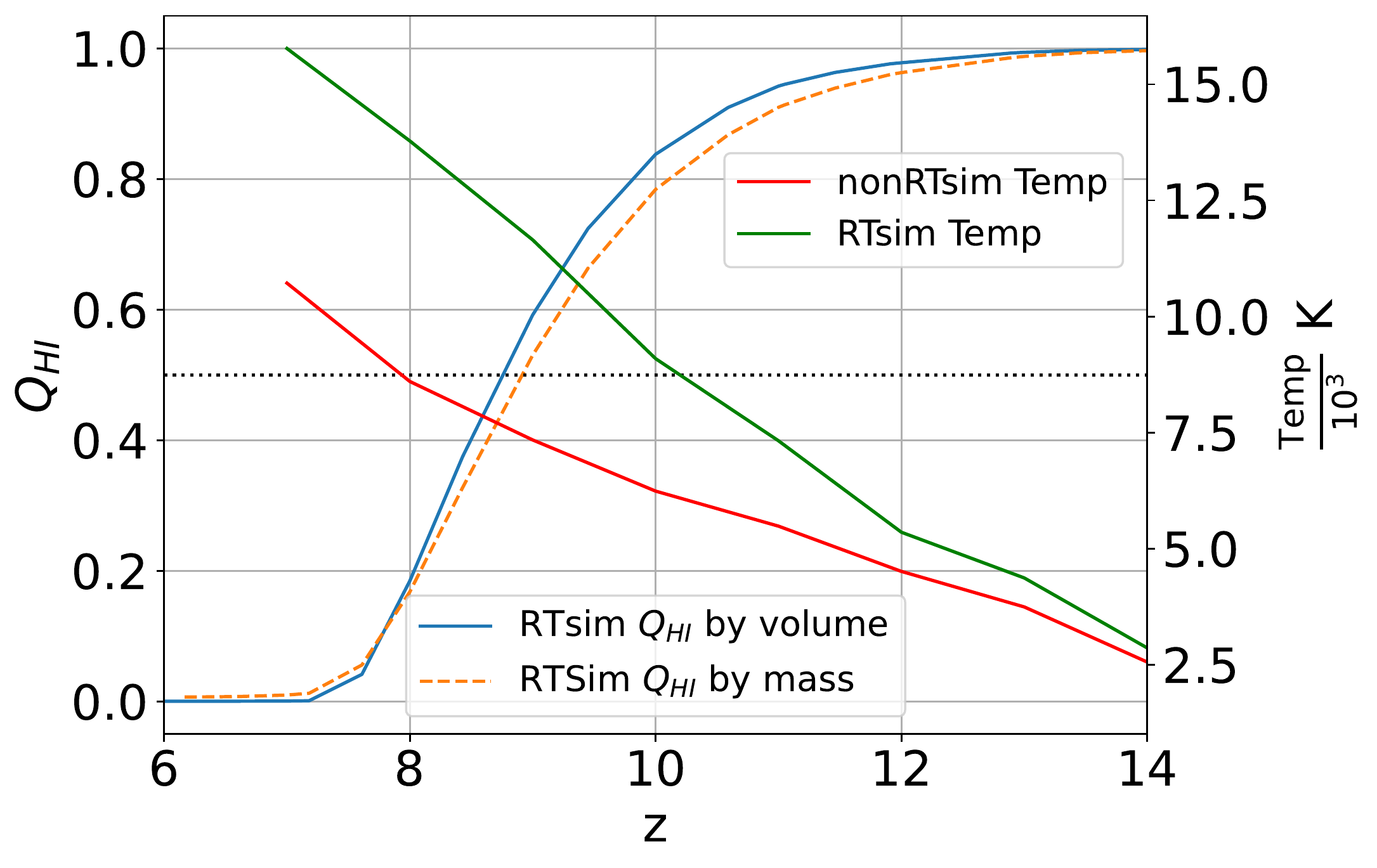}
\caption{The ionization state of the \RTsim{} and the density-weighted temperature of the \nonRTsim{}. The black dotted line indicates 50\% ionization for the \RTsim{}. Just below $z=9$ the \RTsim{} is 50\% reionized while the \nonRTsim{} has reached an average temperature $\gtrapprox$ 7,500 K.}
\label{fig:Q_HI}
\end{center}
\end{figure}

At $z=9$ in the \nonRTsim{}, the \cite{1996ApJ...461...20H} UV background is turned on (the grey region in Fig.~\ref{fig:sfrd}) when the gas has already been heated to an average of $\approx$ 7,500 K, as seen in Fig.~\ref{fig:Q_HI}. At this point, the \RTsim{} is $\approx 50\%$ reionized. Shortly after, by $z=7.5$, this background raises the average gas temperature in the \nonRTsim{} to $\gtrsim$ 9,000 K. This is also the point at which the SFRD in the \RTsim{} begins to overtake the rate in the \nonRTsim{}. This is now possible because of the excess gas in the \RTsim{} as compared to the \nonRTsim{}. The \RTsim{} continues to generate more stars than the \nonRTsim{} for the remainder of the simulation.

\begin{figure*}[t]
\includegraphics[width=.515\textwidth]{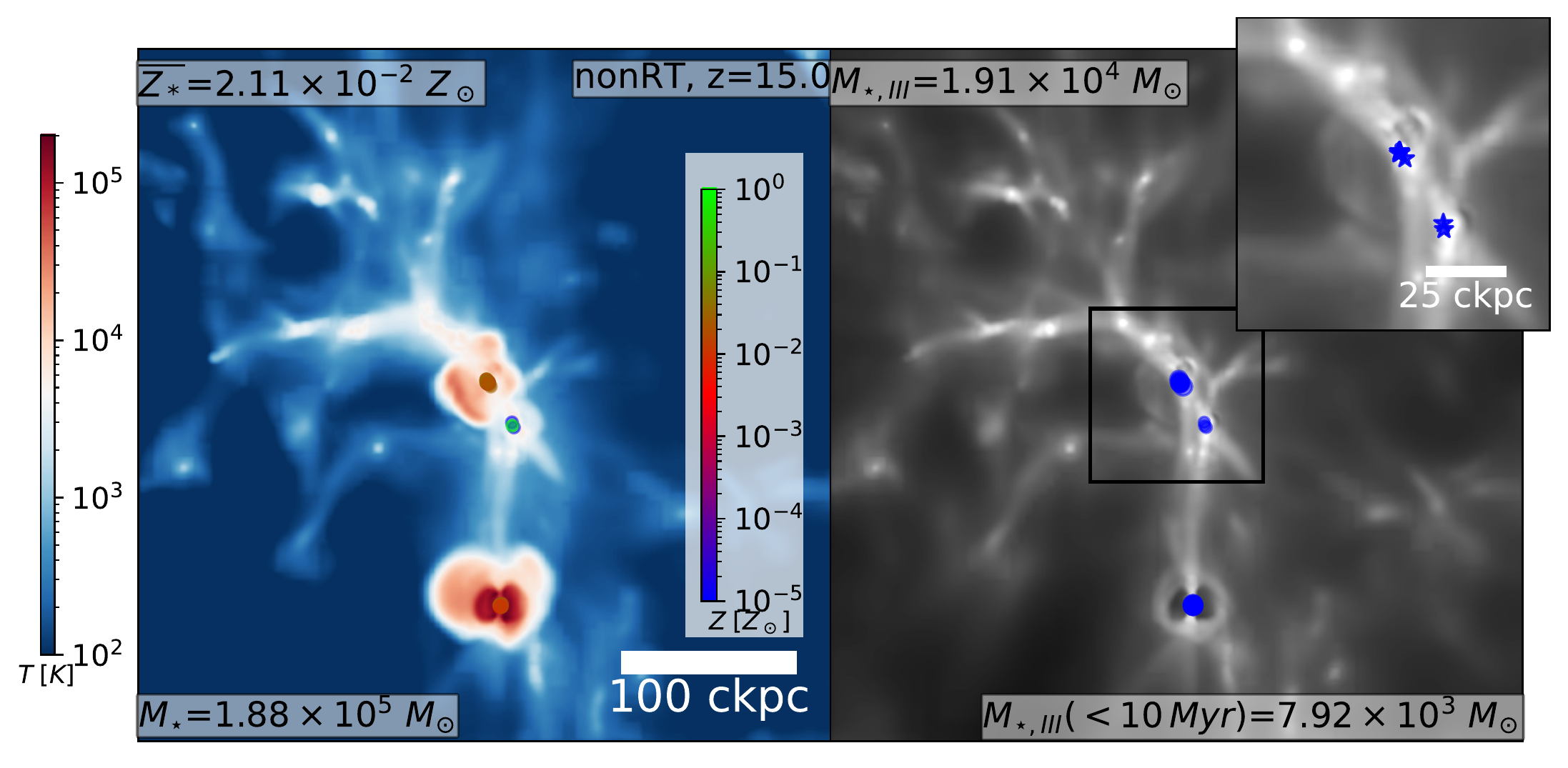} \hspace{-7px}
\includegraphics[width=.48\textwidth]{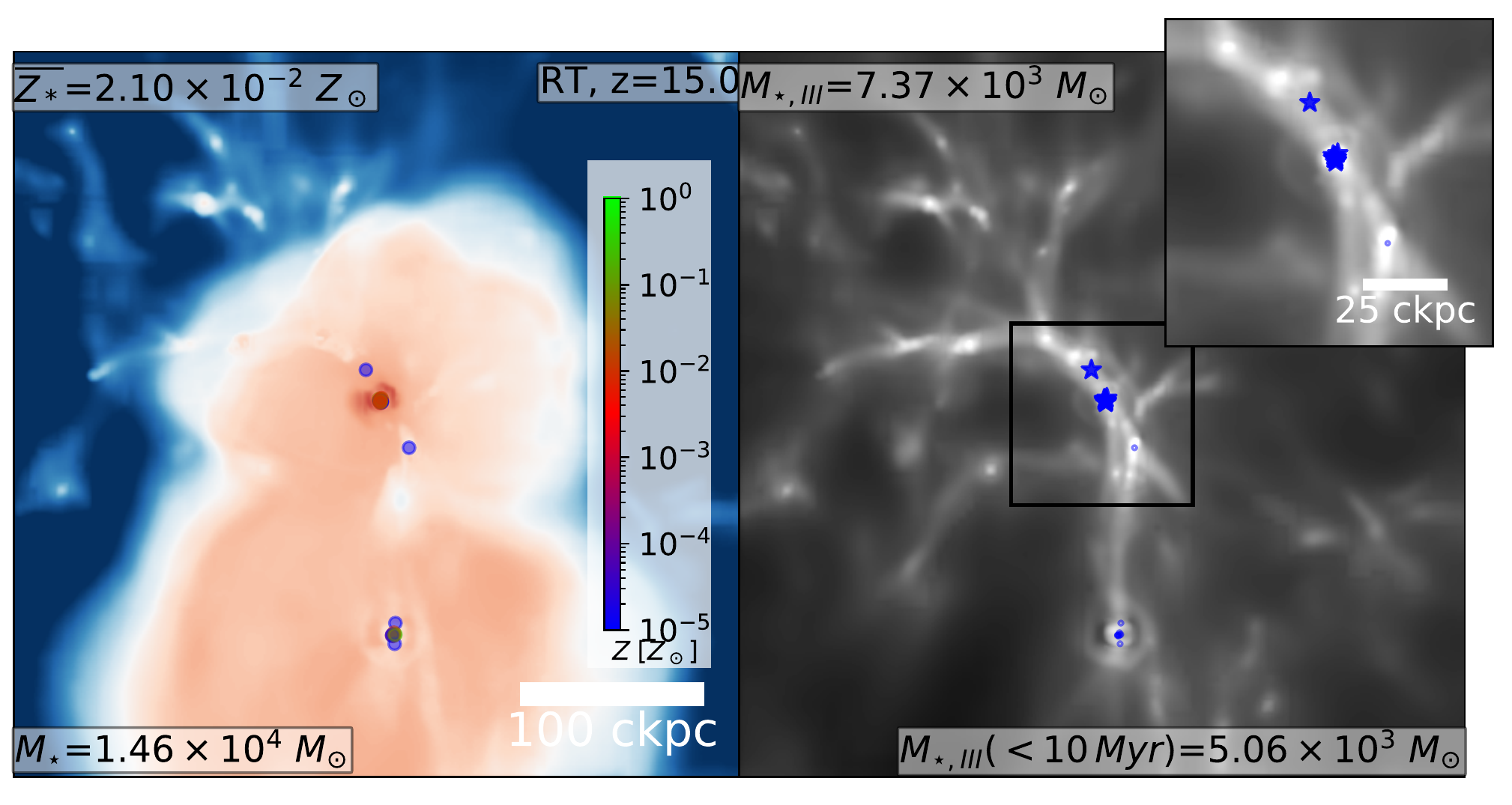} \vspace{-5px} \\
\includegraphics[width=.515\textwidth]{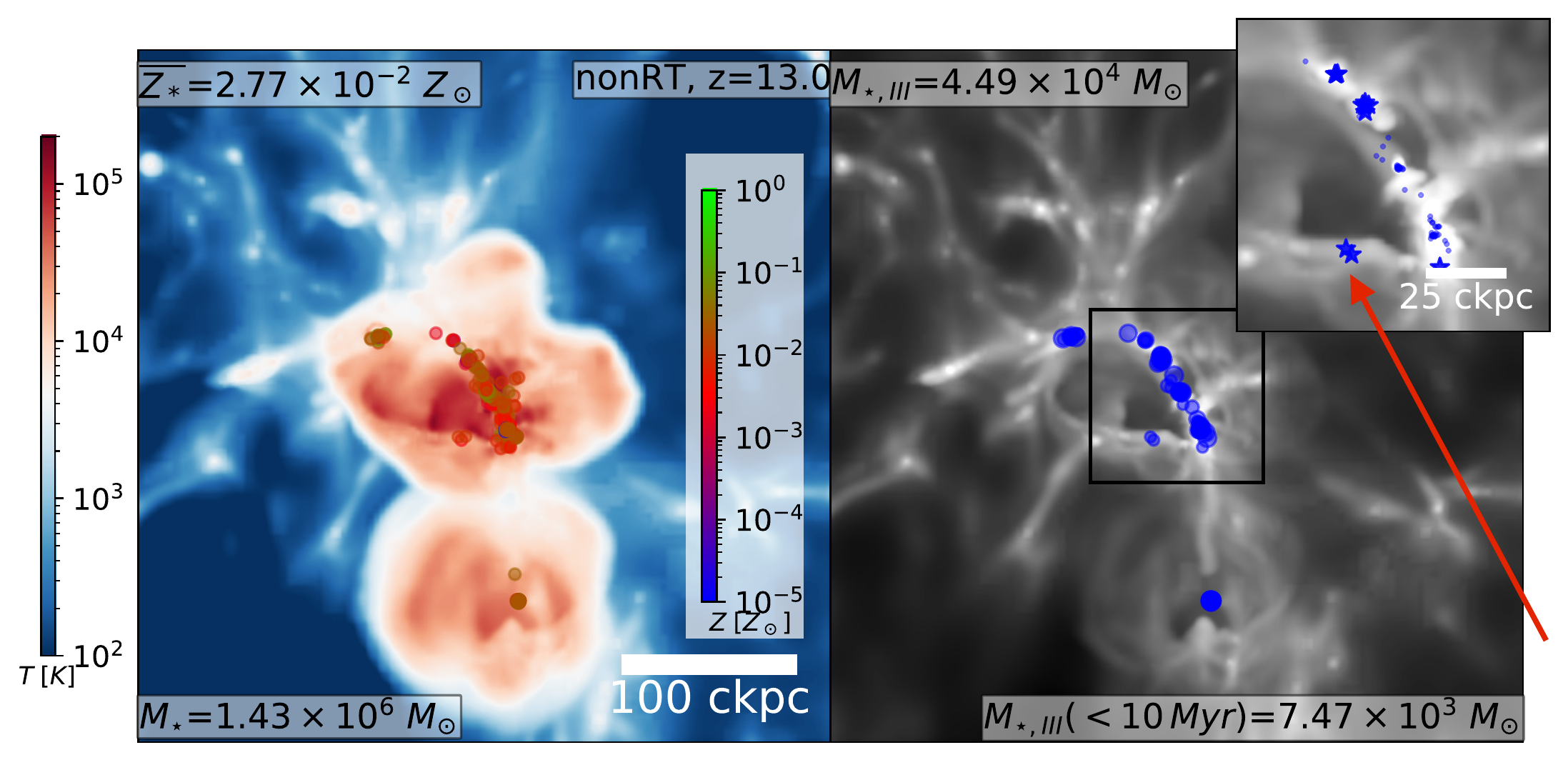} \hspace{-7px}
\includegraphics[width=.48\textwidth]{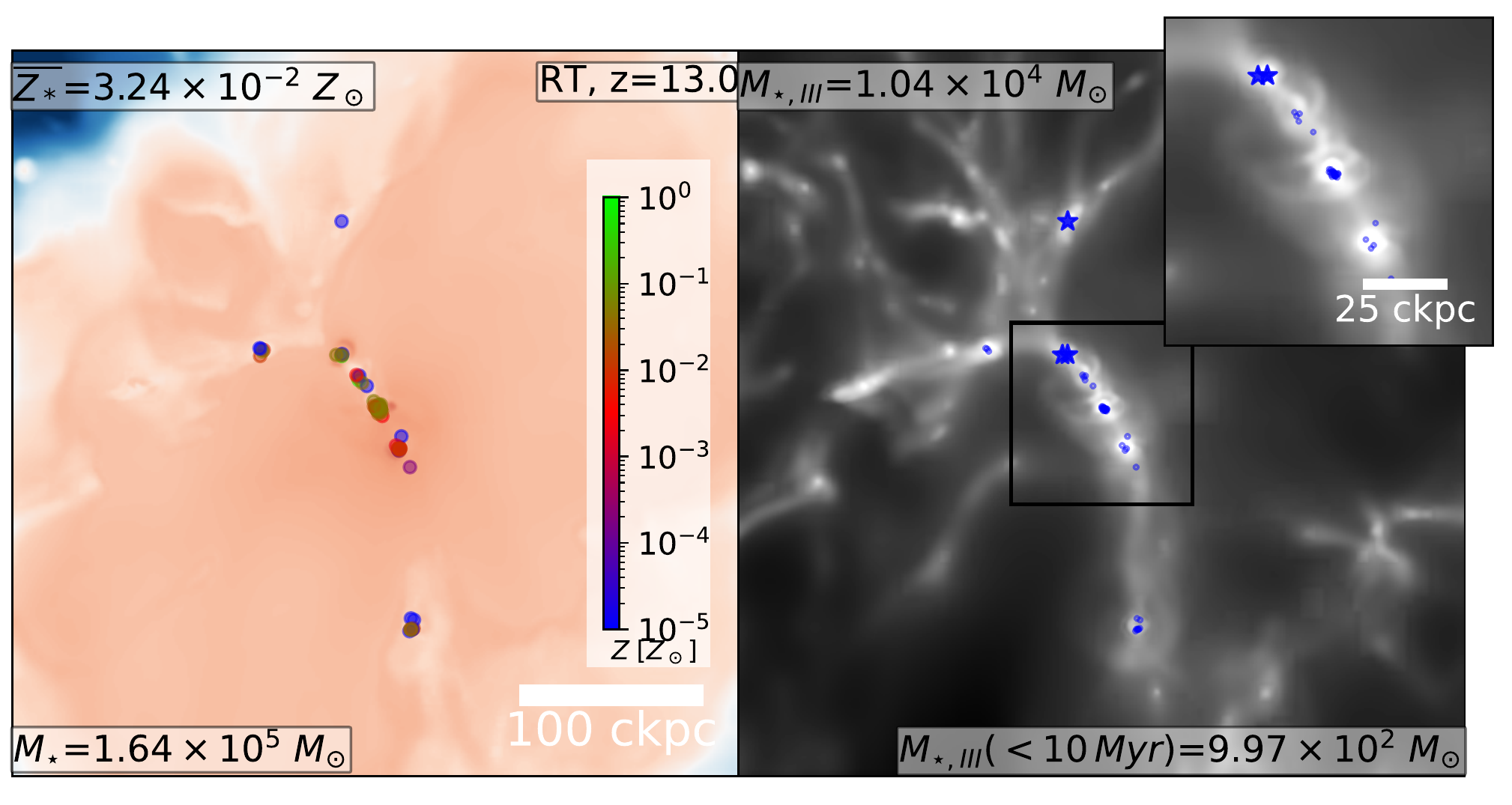} \vspace{-5px} \\
\includegraphics[width=.515\textwidth]{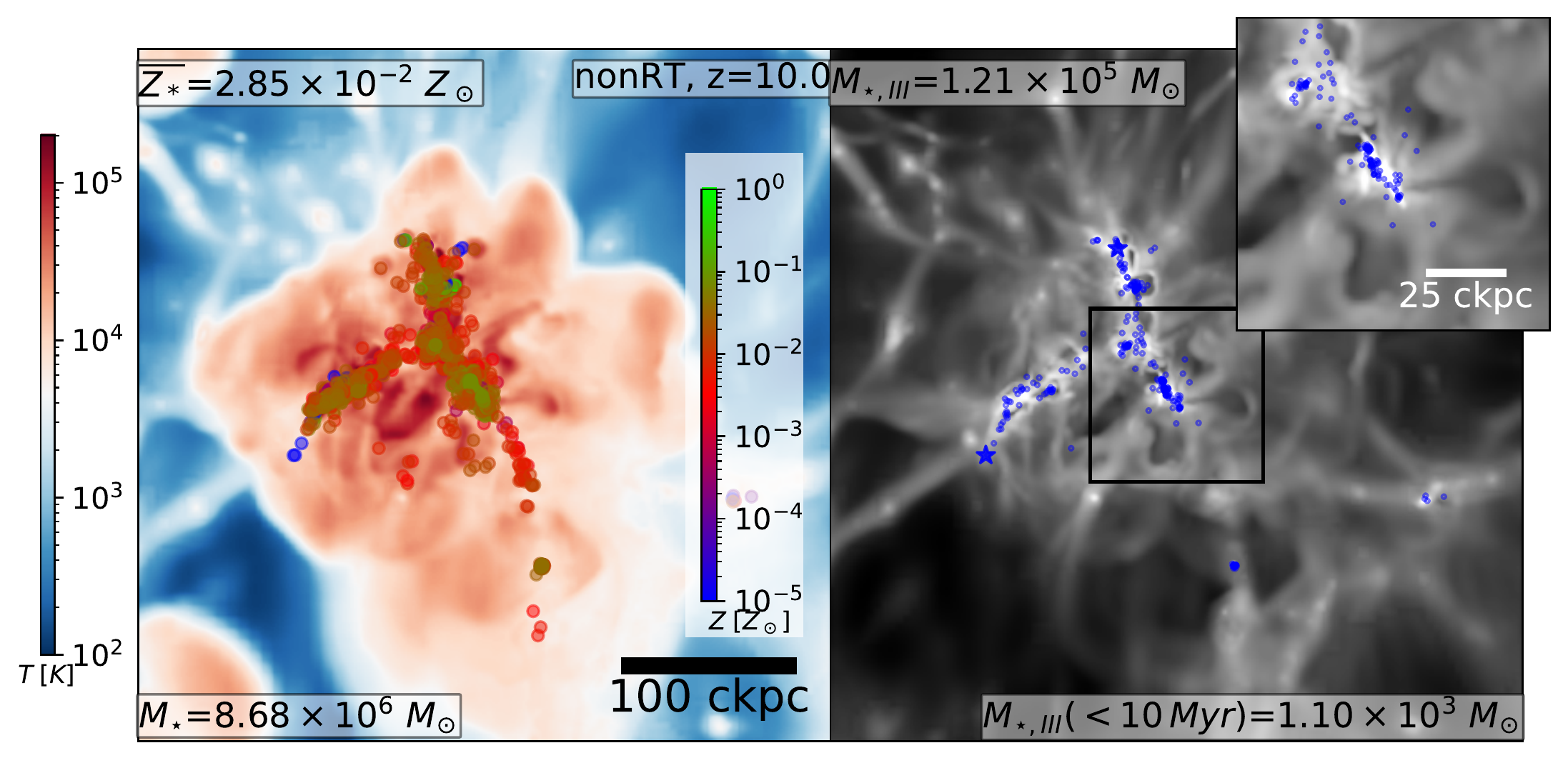} \hspace{-7px}
\includegraphics[width=.48\textwidth]{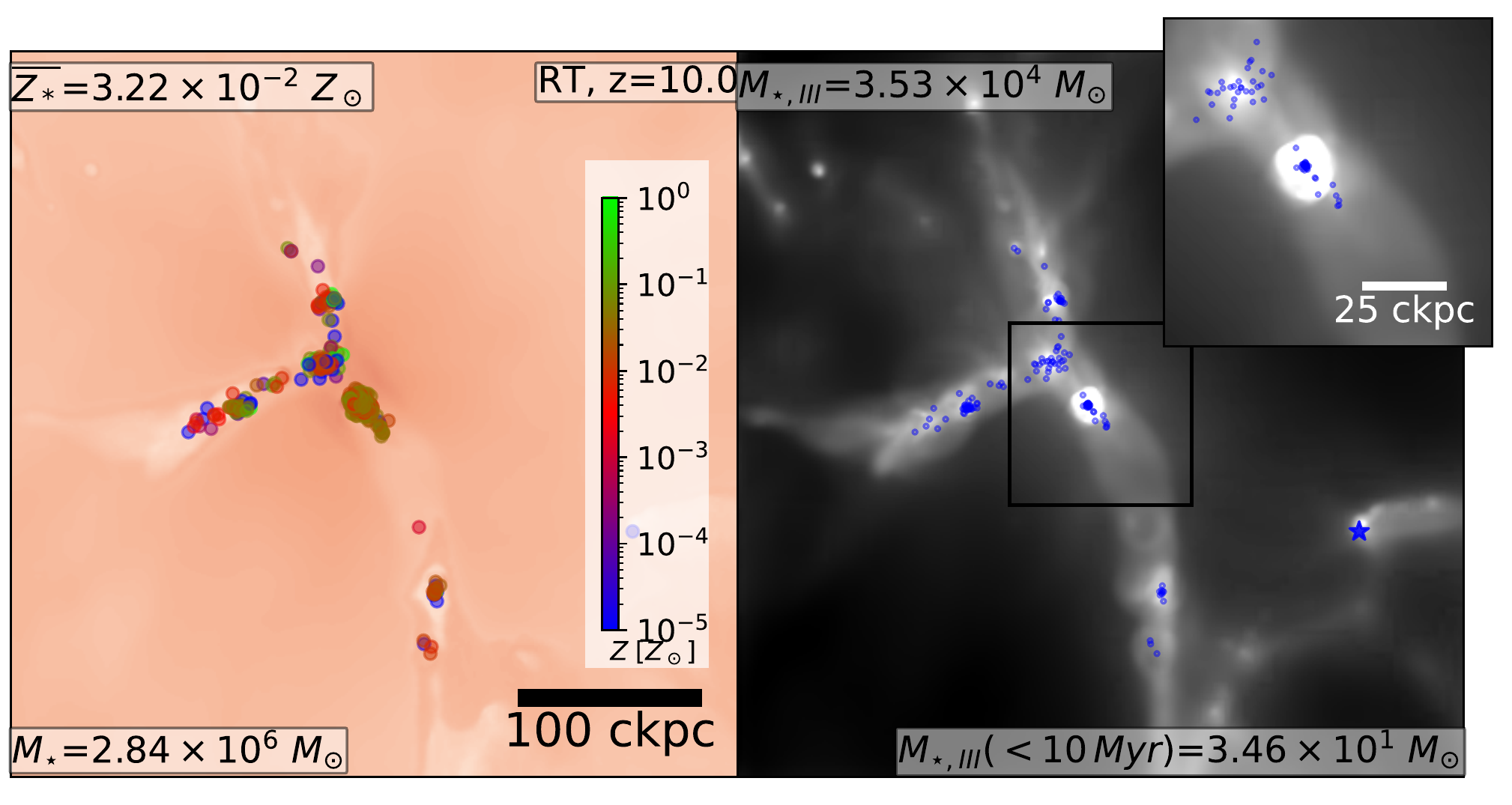} \vspace{-5px} \\
\includegraphics[width=.515\textwidth]{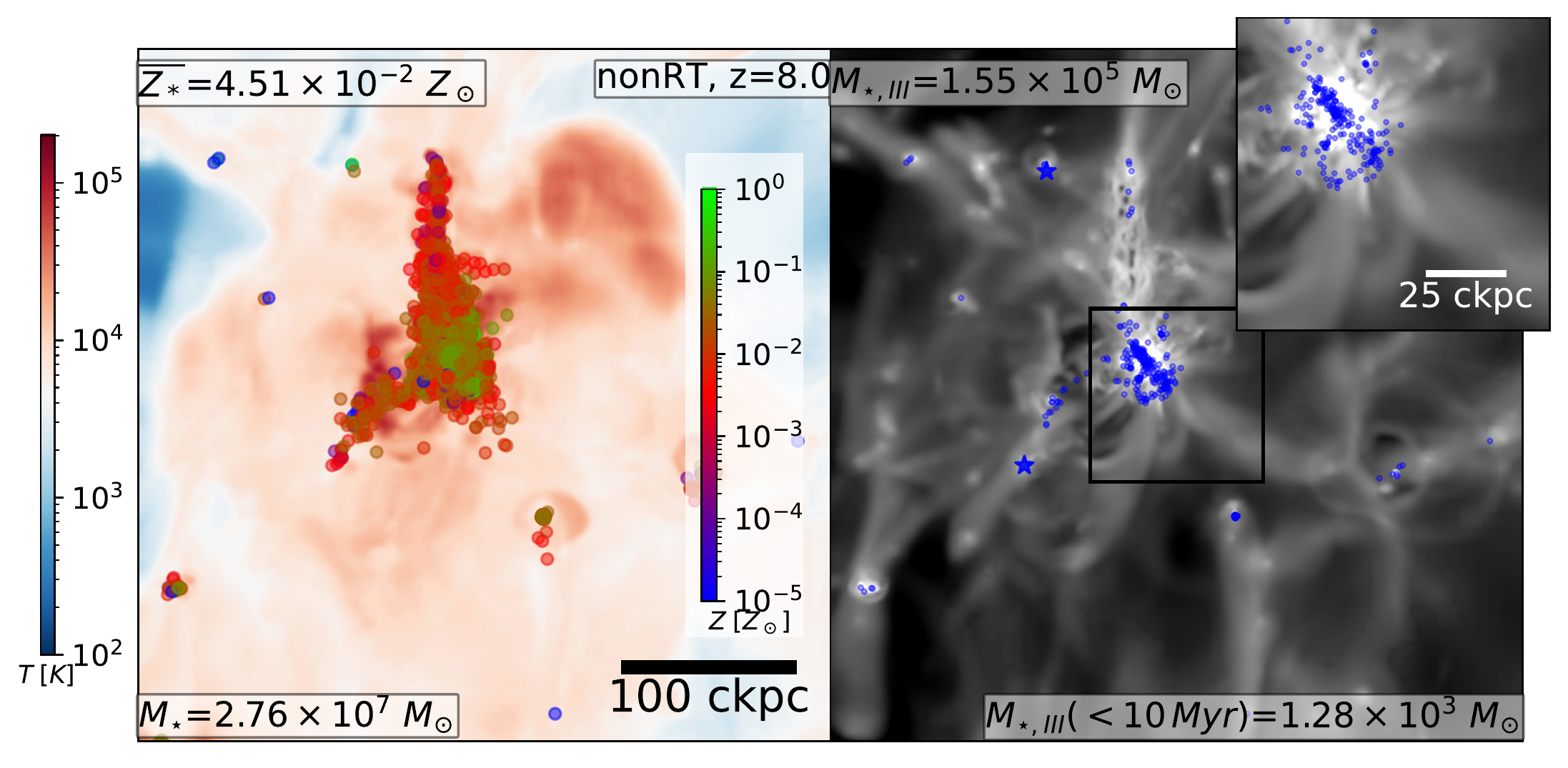} \hspace{-7px}
\includegraphics[width=.48\textwidth]{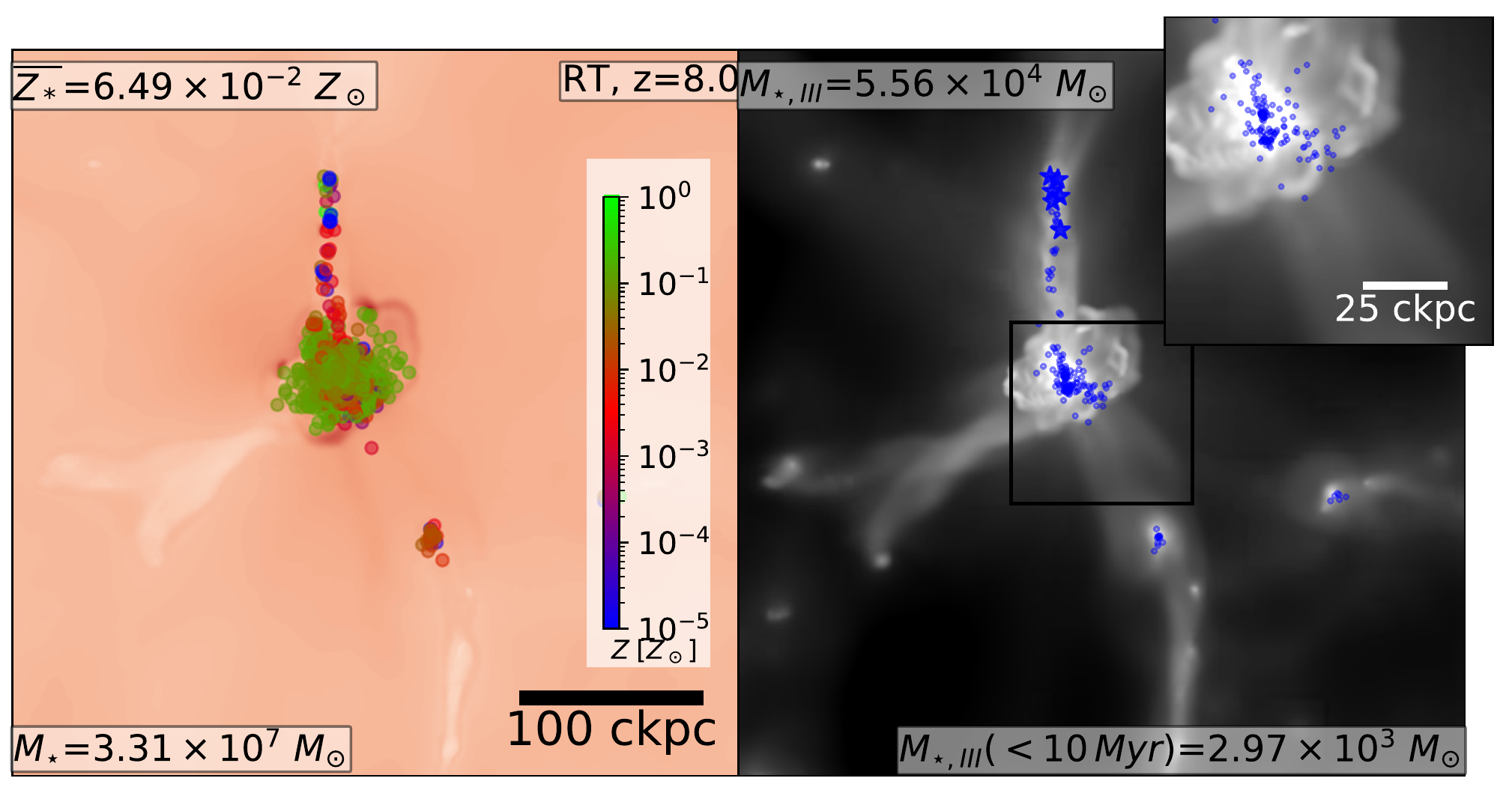} 

\caption{The impact of radiative transfer on the gas, metallicity, and stellar distributions of a representative star-forming region at $z=15$ (top row), $z=13$ (next down), $z=10$ (second from bottom), and $z=8$ (bottom row). Each panel includes a gas temperature plot (left) with SP locations \& metallicity overlaid along with a gas density plot (right; arbitrary units) with Pop III SP (blue dots \& stars) locations overlaid.
The plots depict the density-weighted projection for a 400 ckpc wide region around the earliest star-forming region in the simulations. The inset density plot is a zoom of the central star-forming region with blue circles for Pop III SPs older than 10Myr, and blue stars for Pop III SPs 10 Myr or younger. The left column depicts the \nonRTsim{}, and the right depicts the \RTsim{}.
Note the smoothing effect on the density field caused by radiative feedback. At $z=13$ we see, red arrow, Pop III stars forming in a shock-induced over-density.  This region has been smoothed out in the \RTsim{} by radiative heating. The radiative smoothing of the gas is very apparent at $z=9$ where we see very centralized star formation as compared to the \nonRTsim{}. }
\label{fig:temps}
\end{figure*}

Both simulations generate roughly the same mass in Classical Pop III stars (indicated by  the dotted line in Fig~\ref{fig:sfrd})
at each epoch, down to $z\approx8$. These stars, by definition, are formed in fully pristine gas (cells) unpolluted by SN ejecta \citep{Sarmento2016}. These are areas far enough away from previous star-forming regions that turbulent mixing of ejecta has not begun. These regions are also necessarily separated from existing star-forming regions such that they are not strongly heated by stellar photons. This agreement in the Classical Pop III SFRD between the two simulations would seem to indicate that modeling RT is not crucial to accurately modeling Pop III star formation in the early universe.

However, the overall Pop III SFRD (dashed line in  Fig~\ref{fig:sfrd}) is higher in the \nonRTsim{} than in the \RTsim{}. In fact, over the redshift range $8 \le z \le 18$ it is 4 times higher in the \nonRTsim{}. Thus the excess of Pop III stars in the \nonRTsim{} must be formed in regions of incomplete mixing and the \nonRTsim{} forms more PopIII stars in this region. Since SNe are the primary driver of turbulent mixing, we conclude that radiative heating suppresses Pop III star formation in areas actively stirred by SN energy.

Note that the overall Pop III star formation rate in the \RTsim{} is approximately 10 times that of the Classical Pop III rate demonstrating the need to track Pop III star formation in unresolved cells. By contrast, our earlier 3 Mpc h$^{-1}$ study in \cite{Sarmento2016} displayed an overall-to-Classical Pop III SFRD ratio of $\approx$ 4. The difference is due to the different maximum resolutions used in these studies. In \cite{Sarmento2016} we used a maximum resolution 4 times higher than the one used here. The increased resolution results in a Classical Pop III SFRD that more closely follows the `true' rate since we were able to track smaller gas parcels in that study. Indeed, with infinite resolution there would not be a need to model the subgrid mixing of pollutants since the simulation would effectively track pristine gas clouds down to the star-forming limit. Hence the importance of tracking the unmixed fraction of gas correlates with the resolution of a  simulation: the more coarse the final resolution, the more important following the mixing becomes.

\subsection{Distribution of Pop III Stars}

To illustrate this result in more detail, we examine the gas and SP characteristics centered on a 400 ckpc box around the first region to form stars in Fig.~\ref{fig:temps}.  By $z=15$, about 40 Myr after the start of star formation, the \nonRTsim{} has generated $\approx 10$ times the mass in stars as compared to the \RTsim{}. However, approximately 90\% of the \nonRTsim{}'s SPs in this region are Pop II, indicating ongoing star formation in regions heavily polluted by previous SN. For the \RTsim{}, the Pop II fraction is approximately 50\%, indicating relative suppression in polluted regions.  This again suggests that the \RTsim{} generated fewer stars in the regions closest to the original star-burst where pollution is highest. A high fraction of Pop III stars may aid in the detection of these early galaxies in future surveys \citep{Welch:2022tp,Windhorst2018} possibly via lensing and caustic crossings. While the gas temperatures around existing star-forming regions are comparable between the two simulations, heating is more extensive and likely much more rapid in the \RTsim{}. 

Continuing with Fig.~\ref{fig:temps}, by $z=13$ we see that 97\% of the \nonRTsim{}'s stellar mass is contained in Pop II stars compared to 94\% in the \RTsim{}. The intervening $\approx 62$ Myr between $z=15$ and $z=13$  has enhanced the mixing in this region and we see the \RTsim{} generating many more Pop II stars than during earlier epochs. In fact, the larger number of SN in the \nonRTsim{} has heated the gas in the central region of the \nonRTsim{} such that it is now hotter than the gas in the \RTsim{}. However, there are many more areas of dense gas in the \nonRTsim{} as compared to the \RTsim{}. These appear to be the result of the SN shock-fronts, and they harbor star formation in these partially mixed regions as indicated by the red arrow in the left panel, $z=13$. This `more diffuse' star formation is a characteristic of the \nonRTsim{} as compared to the \RTsim{} as will be further demonstrated.

By $z=10$ both simulations have approximately 99\% of their stellar mass tied up in Pop II stars, while the \nonRTsim{} has $\approx 3$ times the mass in stars as compared to the \RTsim{}.  Overall, we see more star formation away from the central, dense cores in the \nonRTsim{} as a result of star formation in surviving SN-induced over densities. The SN shock fronts are far more well defined in the \nonRTsim{} than in the \RTsim{} and these make more diffuse star formation possible. This effect is most pronounced at $z=10$ and 8. The star-forming regions in both simulations are heated to $\gtrsim 10,000$ K. However, the over densities induced by SN shocks are far more prominent in the \nonRTsim{}, while the gas in the area outside of the main filaments has been effectively smoothed out in the \RTsim{}. Furthermore, SPs in the \RTsim{} are more closely align with the dense filaments than they are in the \nonRTsim{} -- again evidence that areas of gas compressed by SN shocks provide regions for star formation. This is hard to determine quantitatively from these plots. However the importance of accounting for radiative-transfer effects when modeling the morphology of high redshift galaxies is exemplified by this difference between the simulations.

\subsection{The 2-point Correlation Function in Time and Space}

To more directly quantify this effect, we turn to a measure of the spatial and temporal distribution of star-forming regions. The mass-weighted two-point spatial correlation function, $\xi_2(r)$, is defined as the excess probability that two star particles are separated by a distance $r$ relative to a uniform distribution.  It can be computed from the star particles in our simulation as
\begin{equation}
\xi_2(r_k) =  \frac{\sum_{i,j \in k} m^*_i m^*_j}{M^*_{\rm tot} \rho^* \Delta V_k} - 1,
\label{eqn:scf}
\end{equation}
where $m^*_i$ and $m^*_j$ are the masses of star particles $i$ and $j$, $M^*_{\rm tot}$ and $\rho^*$ the total mass and mass density of stars in the simulation, and the sum is over all pairs $i$ and $j$ such that the separation of the particles is within a bin $k$, centered on a distance $r_k$ and with a corresponding volume  $\Delta V_k = \frac{4 \pi}{3}(r_{k+1/2}^3-r_{k-1/2}^3),$ where $r_{k-1/2}$ and $r_{k+1/2}$ are the inner and outer boundaries of the bin.

Similarly, we can define the two-point temporal-spatial correlation function, $\xi'_2(r_k,t_l),$  as the excess probability that the positions and formation times of star particles are separated by a distance $r$ and time $t$ relative to a uniform distribution (in both time and space).  This can be computed as 
\begin{equation}
\xi'_2(r_k,t_l) =  \frac{\sum_{i,j \in k,l} m^*_i m^*_j}{M^*_{\rm tot} \dot \rho^* \Delta V_k E_l} - 1,
\label{eqn:tscf}
\end{equation}
where the sum is now over all pairs $i$ and $j$ such that the separation of the particles is within a bin $k$ and the difference in formation times between the two particles is within a bin $l$, centered on $t_l$ and with a corresponding difference in birth times within a bin $E_l = 2 (t_{l+1/2}-t_{l-1/2})$, where  $t_{l-1/2}$ and $t_{l+1/2}$ define the boundaries of temporal bin and the factor of $2$ accounts for the fact that the bin is defined to be the absolute value of the difference of the birth times of the two star particles.

\begin{figure*}[t]
\begin{center}
\includegraphics[width=0.529\textwidth]{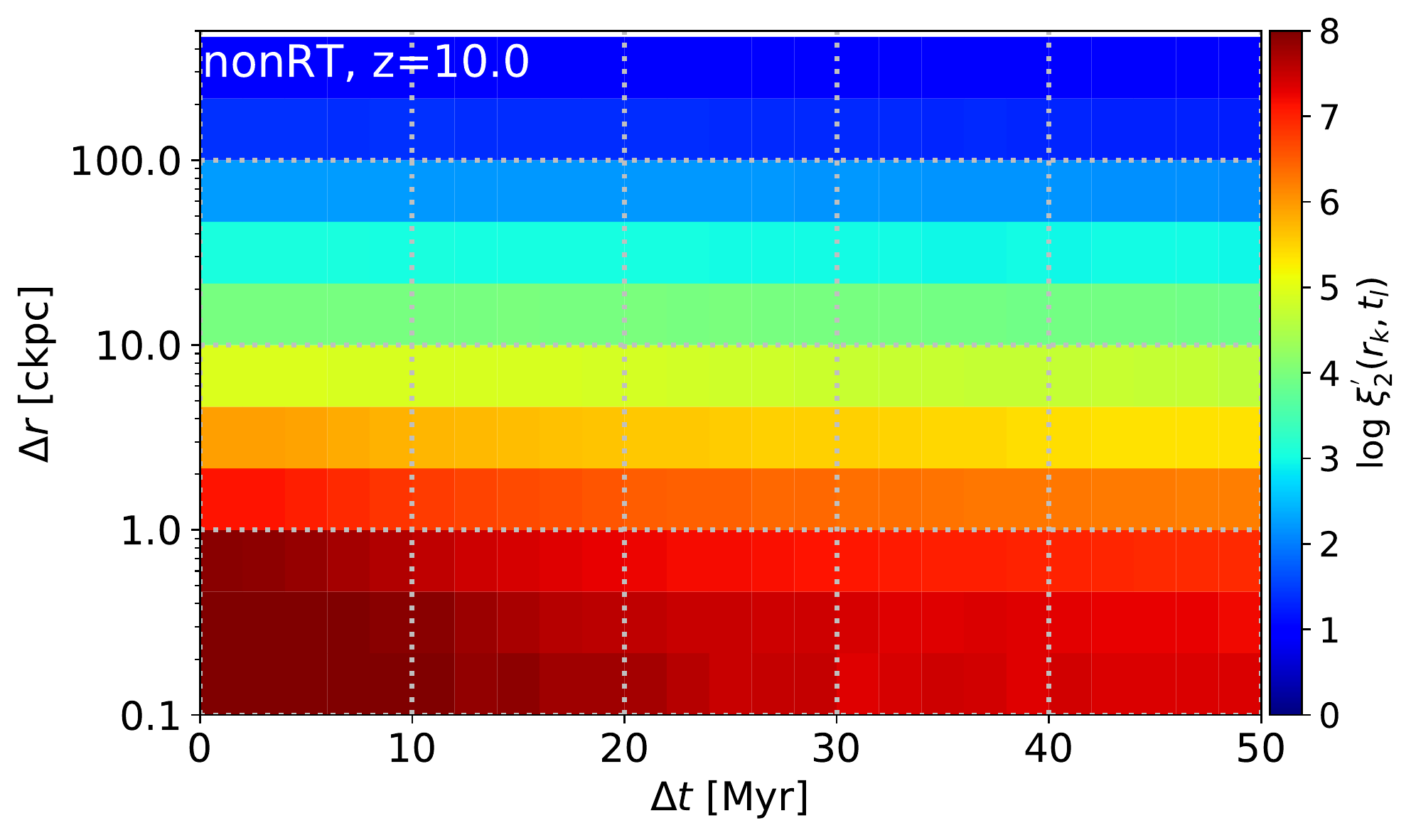} \hspace{-5px}
\includegraphics[width=0.468\textwidth]{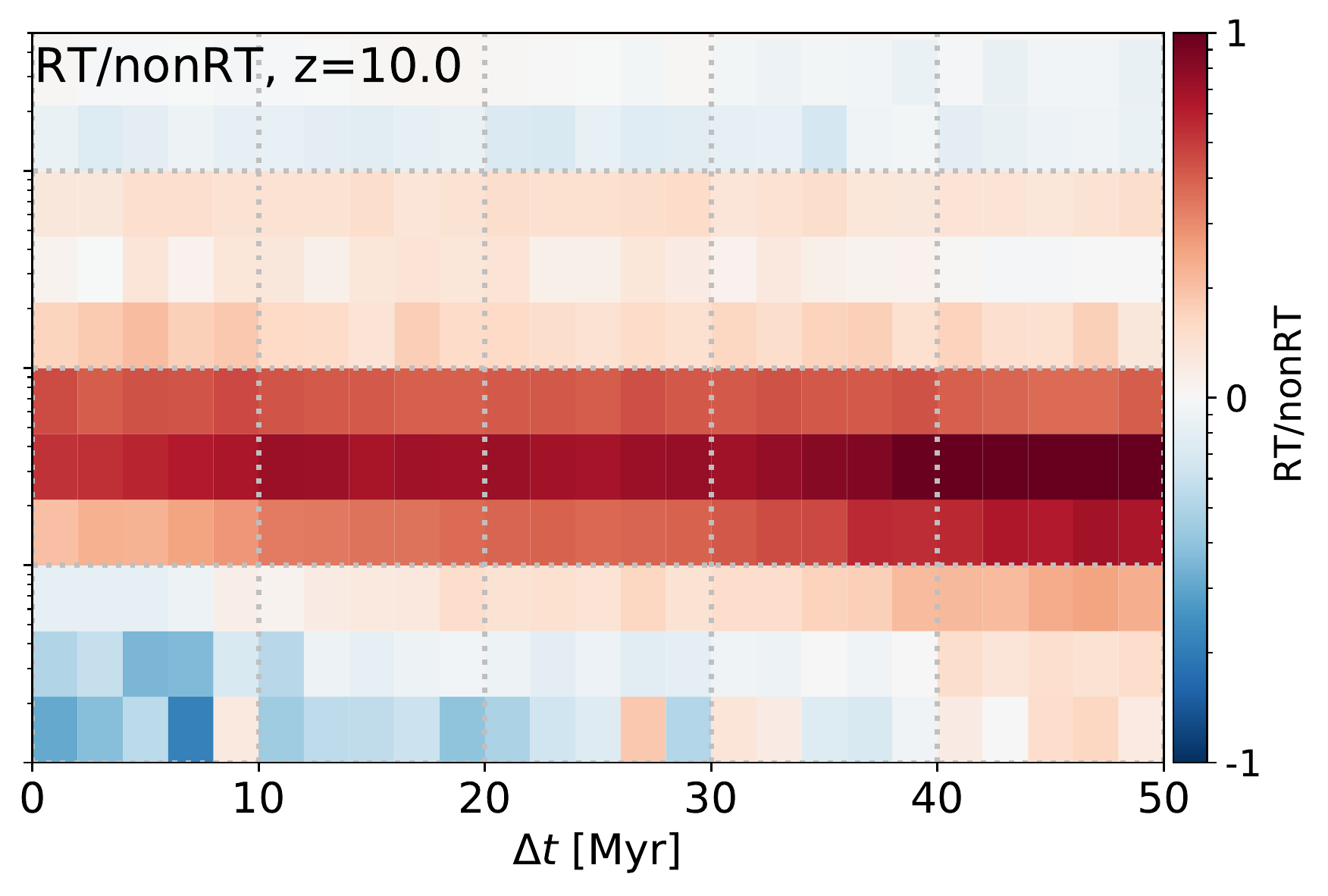} \\
\includegraphics[width=0.529\textwidth]{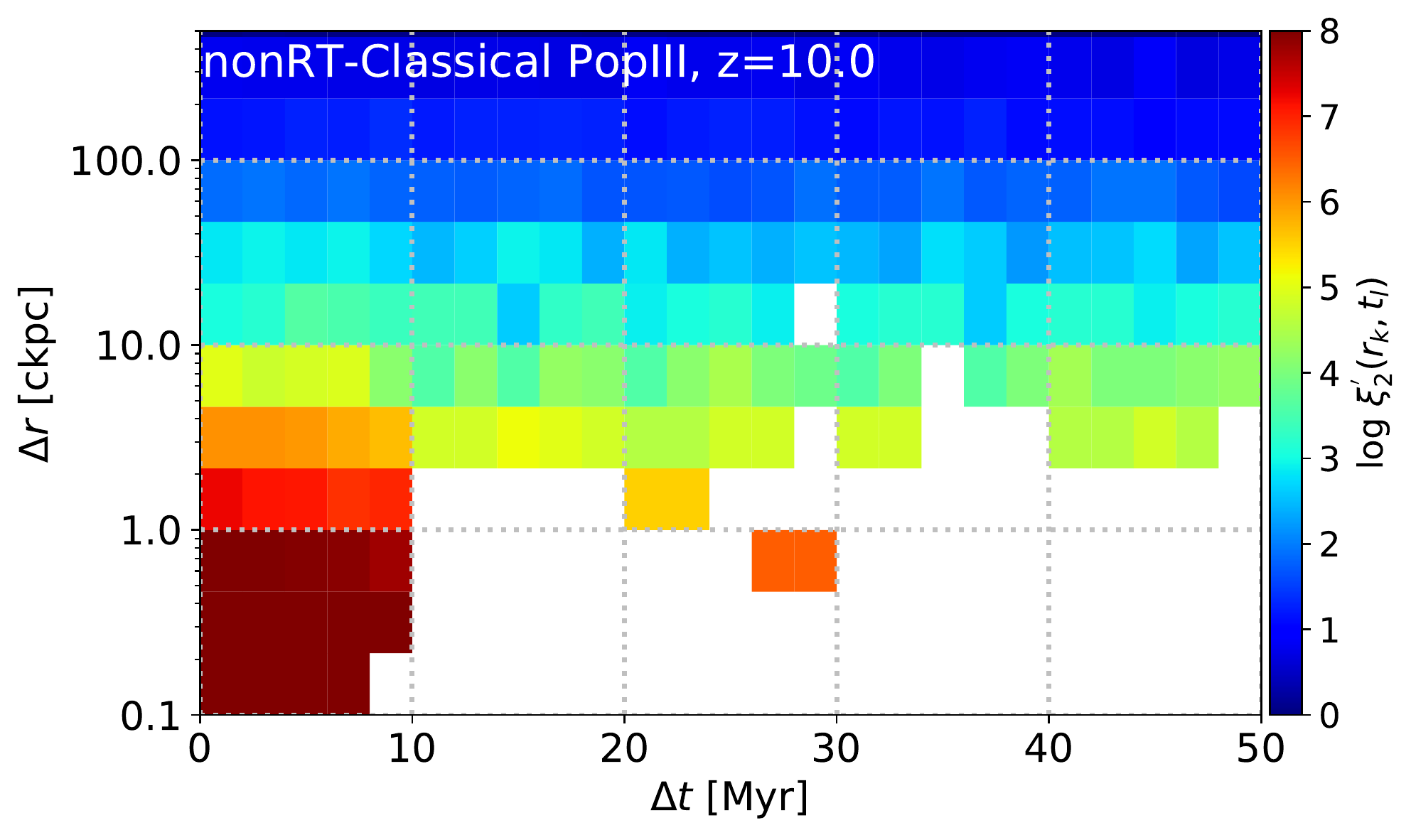}\hspace{-7px}
\includegraphics[width=0.468\textwidth]{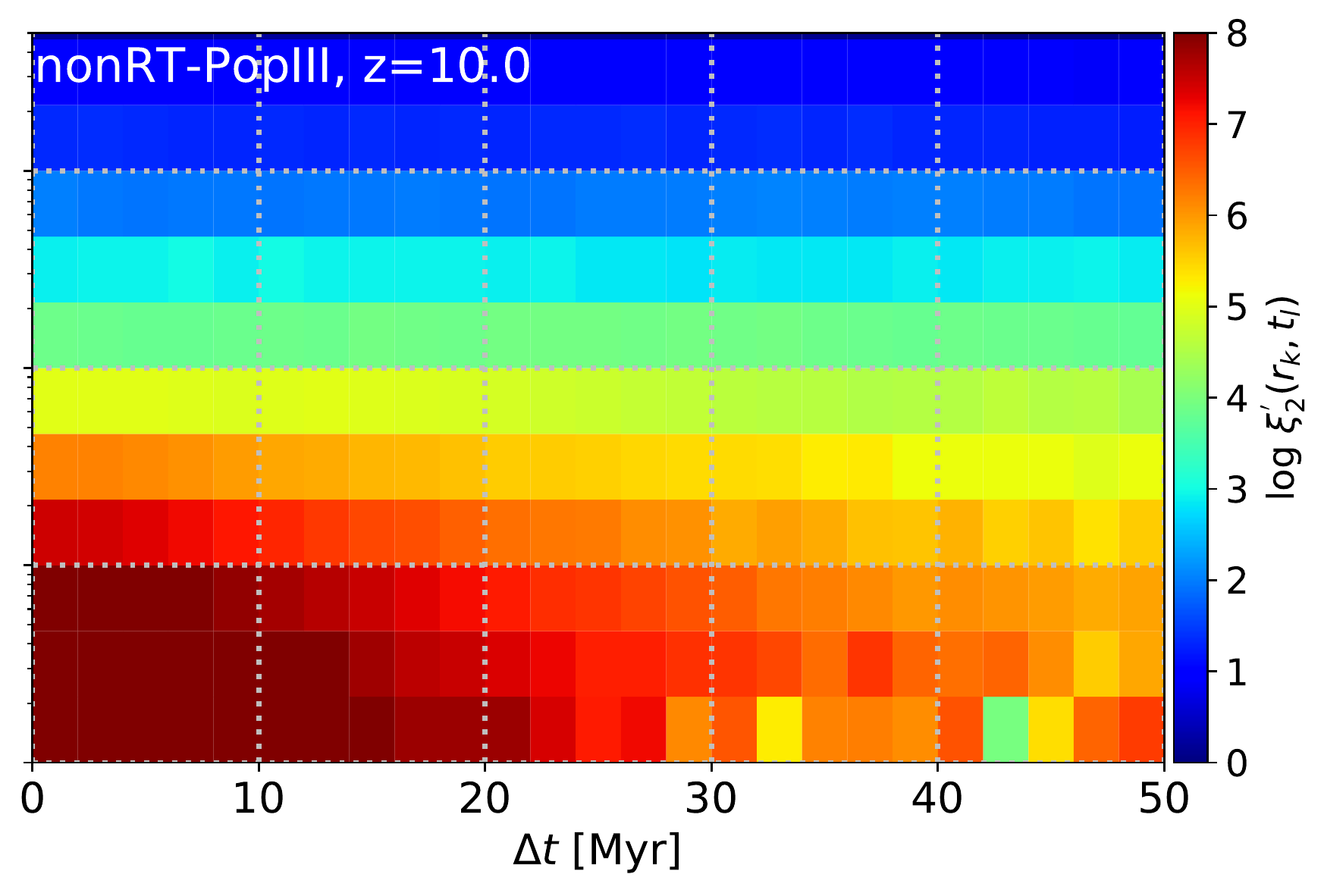}
\caption{The excess probability, $\xi_2^{'}(r_k,t_l)$, of finding pairs of SPs at a given spatial separation, $y$-axis, and a given separation in birth-time, $x$-axis, at $z=10$. Temporal bins are 2 Myrs wide, spatial are 0.33 dex. 
{\em Upper Left:} $\xi_2^{'}(r_k,t_l)$ for the \nonRTsim{}, which shows $\approx$ 20 Myr, bursts of star formation followed by a gradual decline to $\Delta t = 0 \rightarrow 50$, on scales of $\Delta r \lesssim 1$ ckpc.  
{\em Upper Right:}  The ratio of $\xi_2^{'}(r_k,t_l)$ in the \RTsim{} and \nonRTsim{},  which indicates a relative suppression of star formation in the \RTsim{} galaxy cores at $\Delta r < 1$ ckpc and out to $\approx 30$ Myr, as well as an enhanced probability of finding SPs in the \RTsim{} at  distances of 1-10 ckpc.
{\em Lower Left:} The excess probability, $\xi_2^{'}(r_k,t_l)$, in the nonRT sim, computed for the Classical Pop III stars, which  can only form in simulation cells composed solely of unpolluted gas. This shows a near-complete lack of Classical Pop III star formation in regions $\Delta t > 10$ Myr and $\Delta r < 1$ due to local pollution by SN ejecta. 
{\em Lower Right:} $\xi_2^{'}(r_k,t_l)$ computed for all Pop III stars in the \nonRTsim{}, including those formed in areas of incomplete mixing as identified by our subgrid model.   This falls off smoothly both spatially and temporally, demonstrating the importance of modeling metal mixing.}
\label{fig:popIIIyng}
\end{center}
\end{figure*}

\begin{figure*}[t]
\begin{center}
\includegraphics[width=0.245\textwidth]{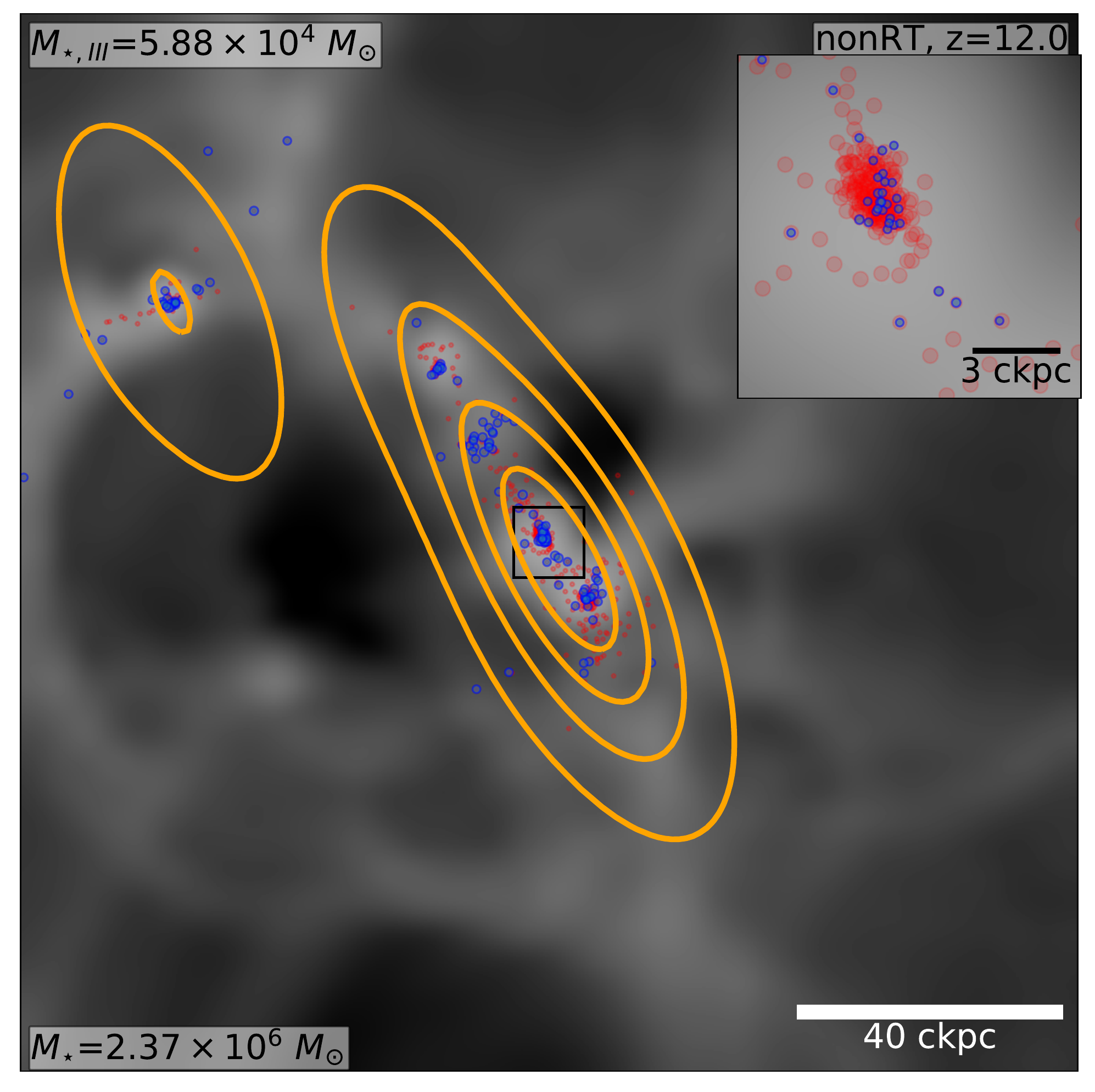} \hspace{-8px}
\includegraphics[width=0.245\textwidth]{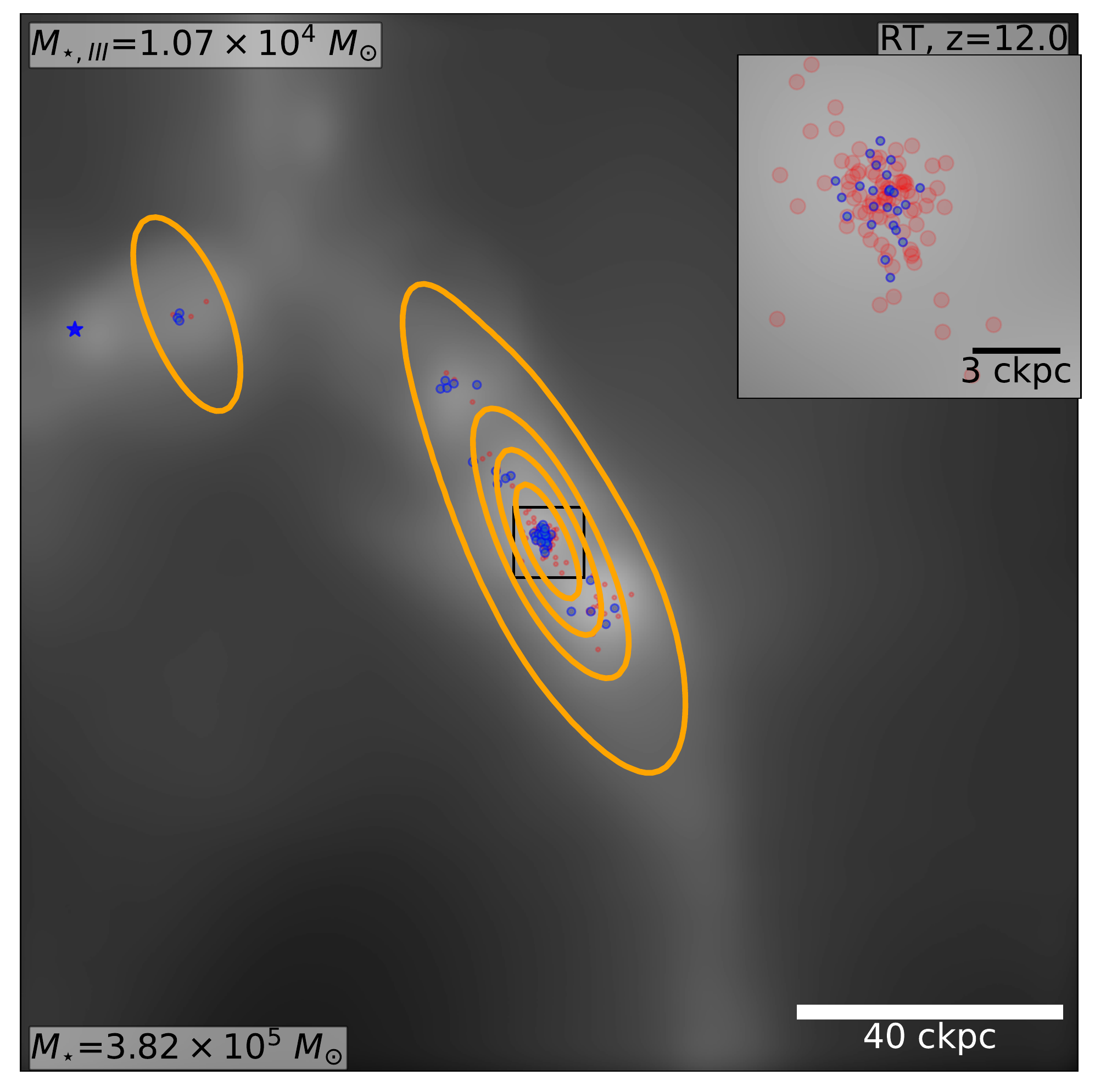} 
\includegraphics[width=0.245\textwidth]{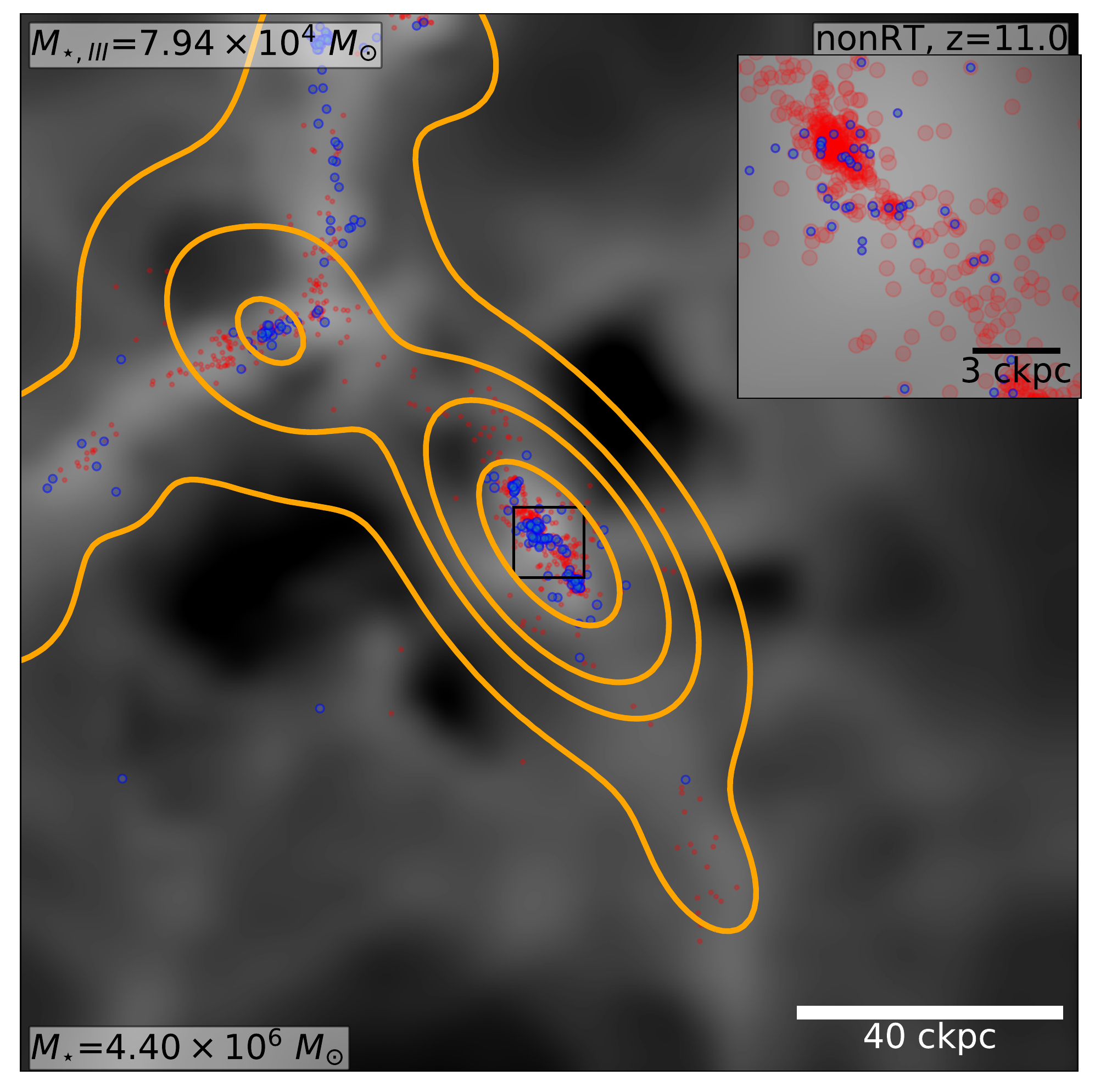} \hspace{-8px}
\includegraphics[width=0.245\textwidth]{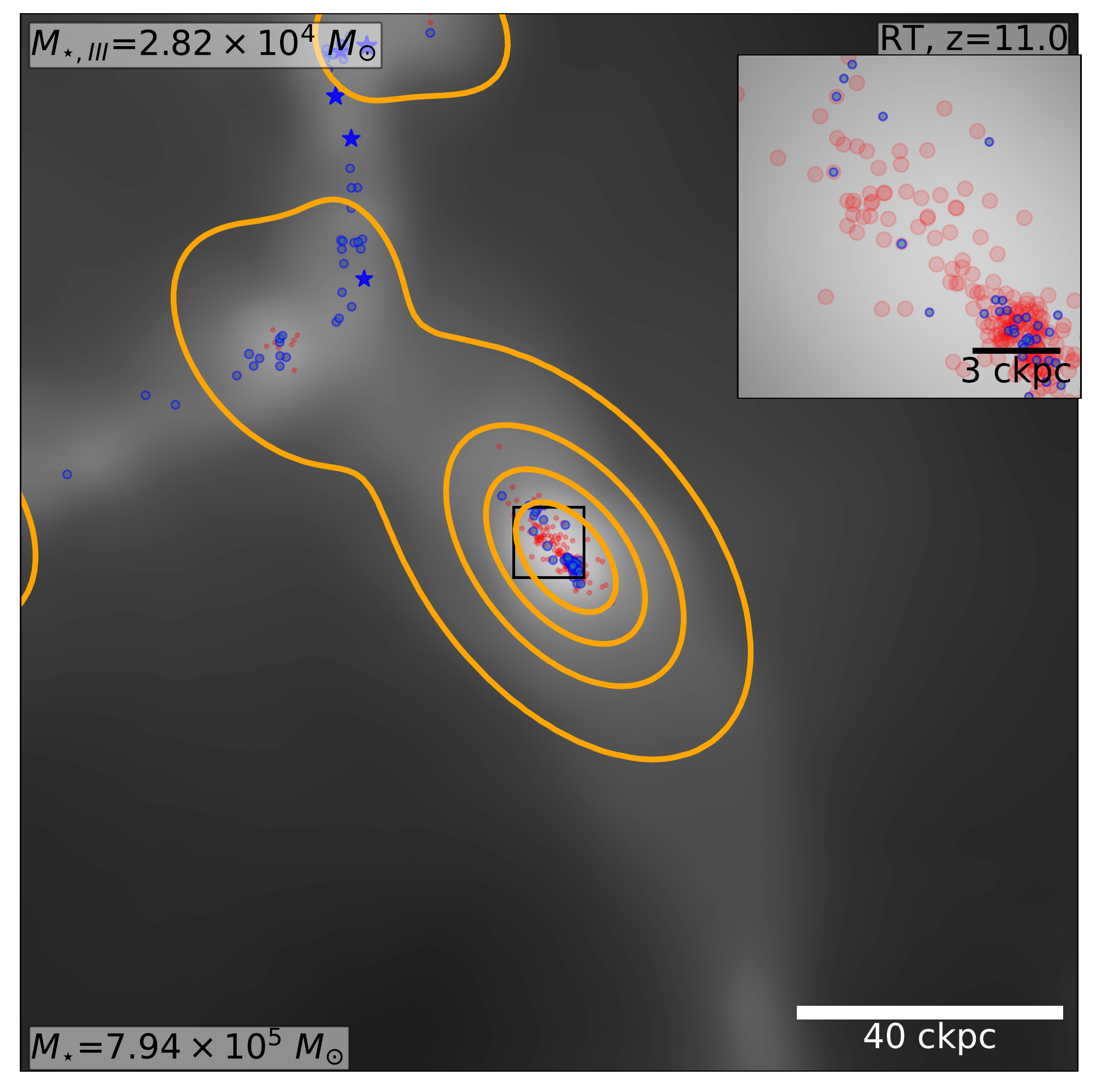} \\
\includegraphics[width=0.245\textwidth]{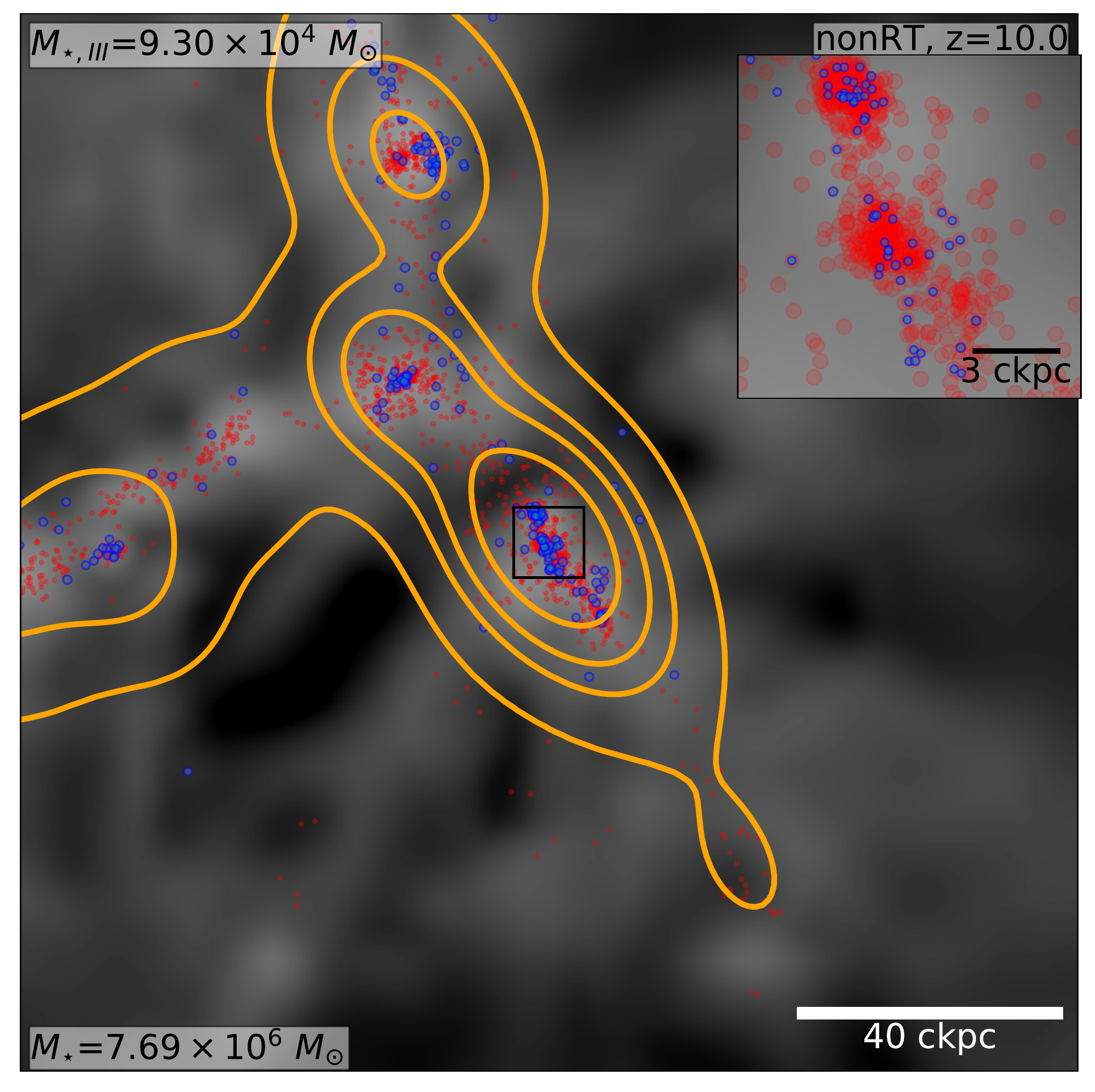} \hspace{-8px}
\includegraphics[width=0.245\textwidth]{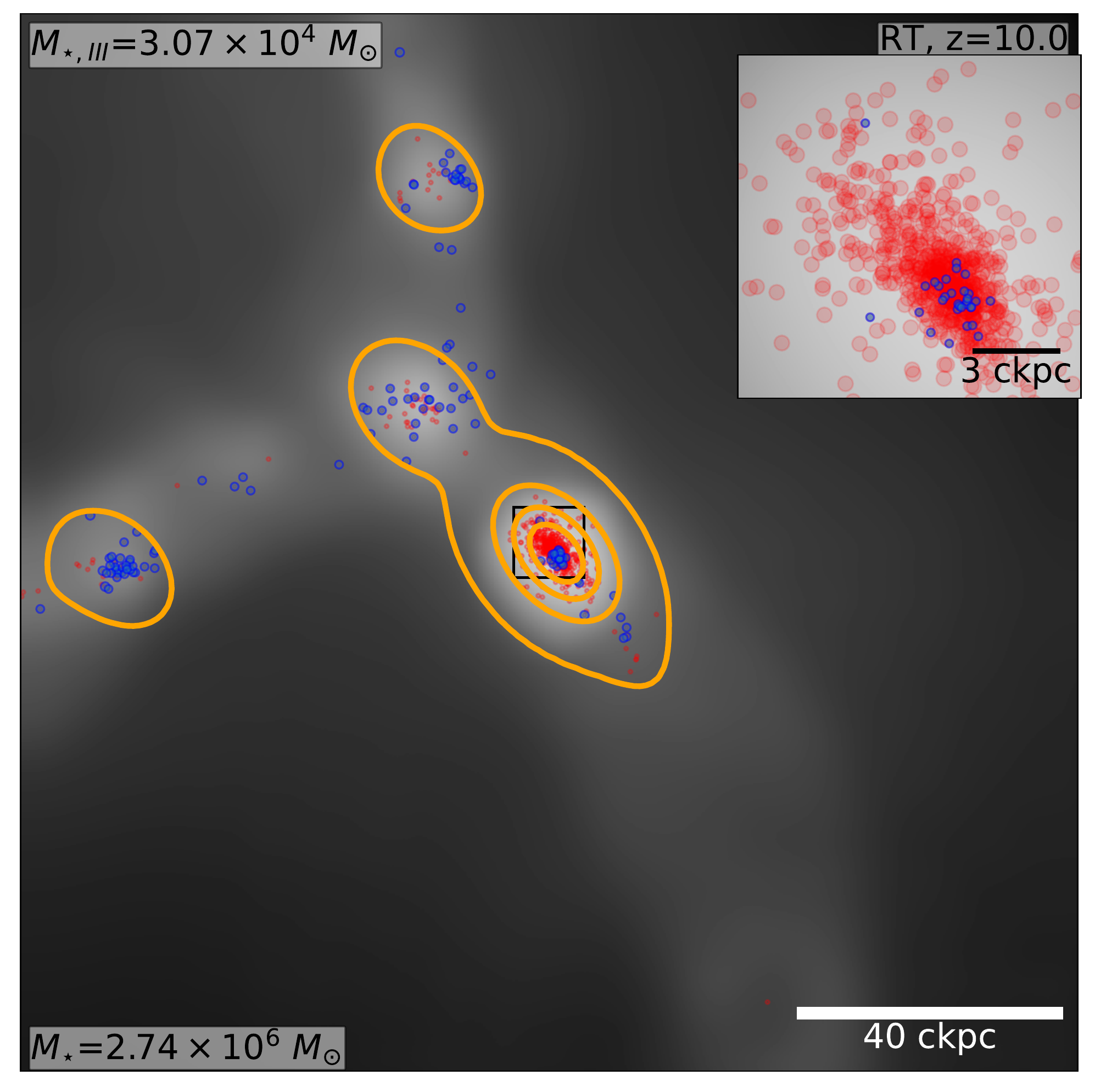}
\includegraphics[width=0.245\textwidth]{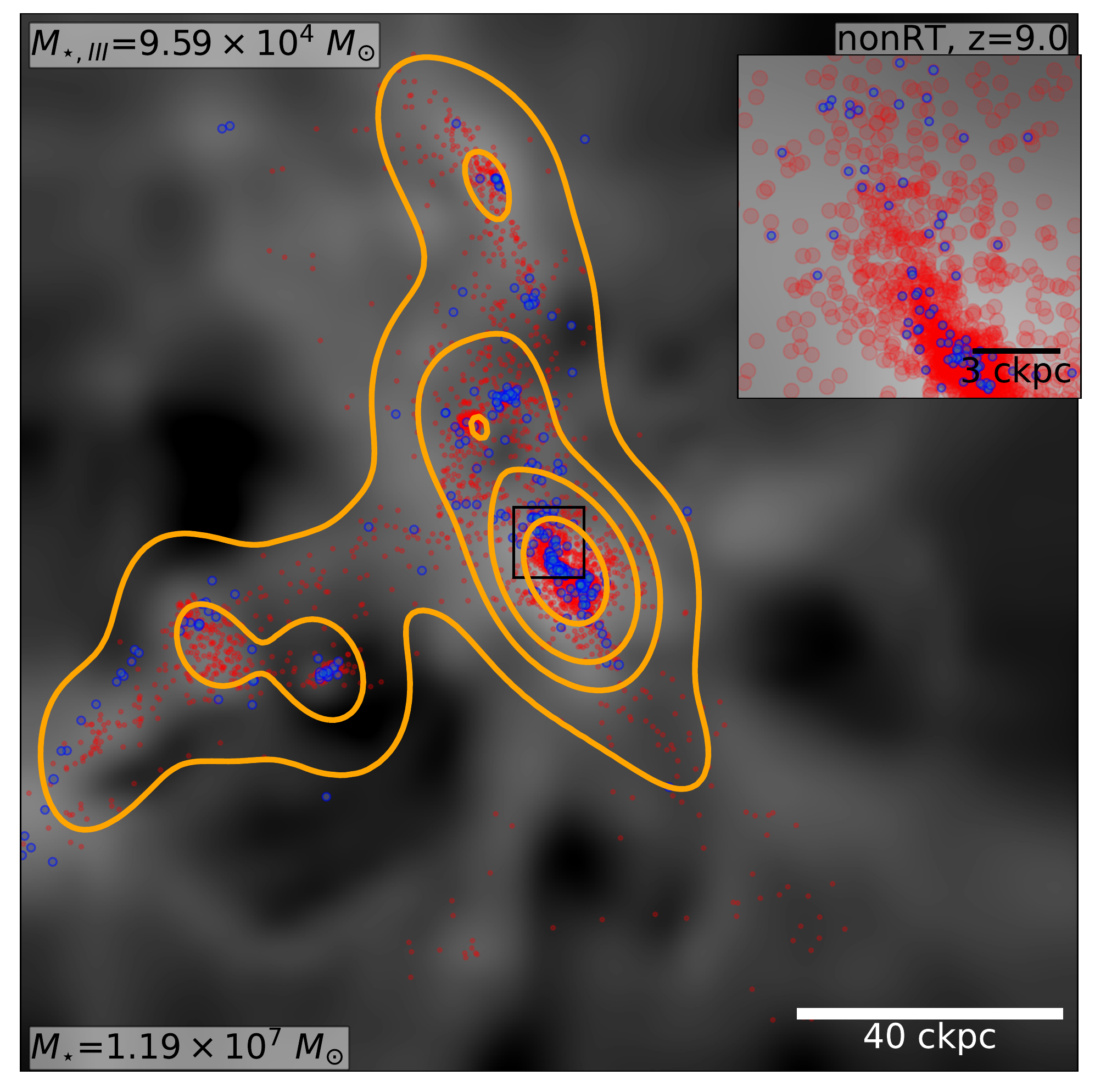} \hspace{-8px}
\includegraphics[width=0.245\textwidth]{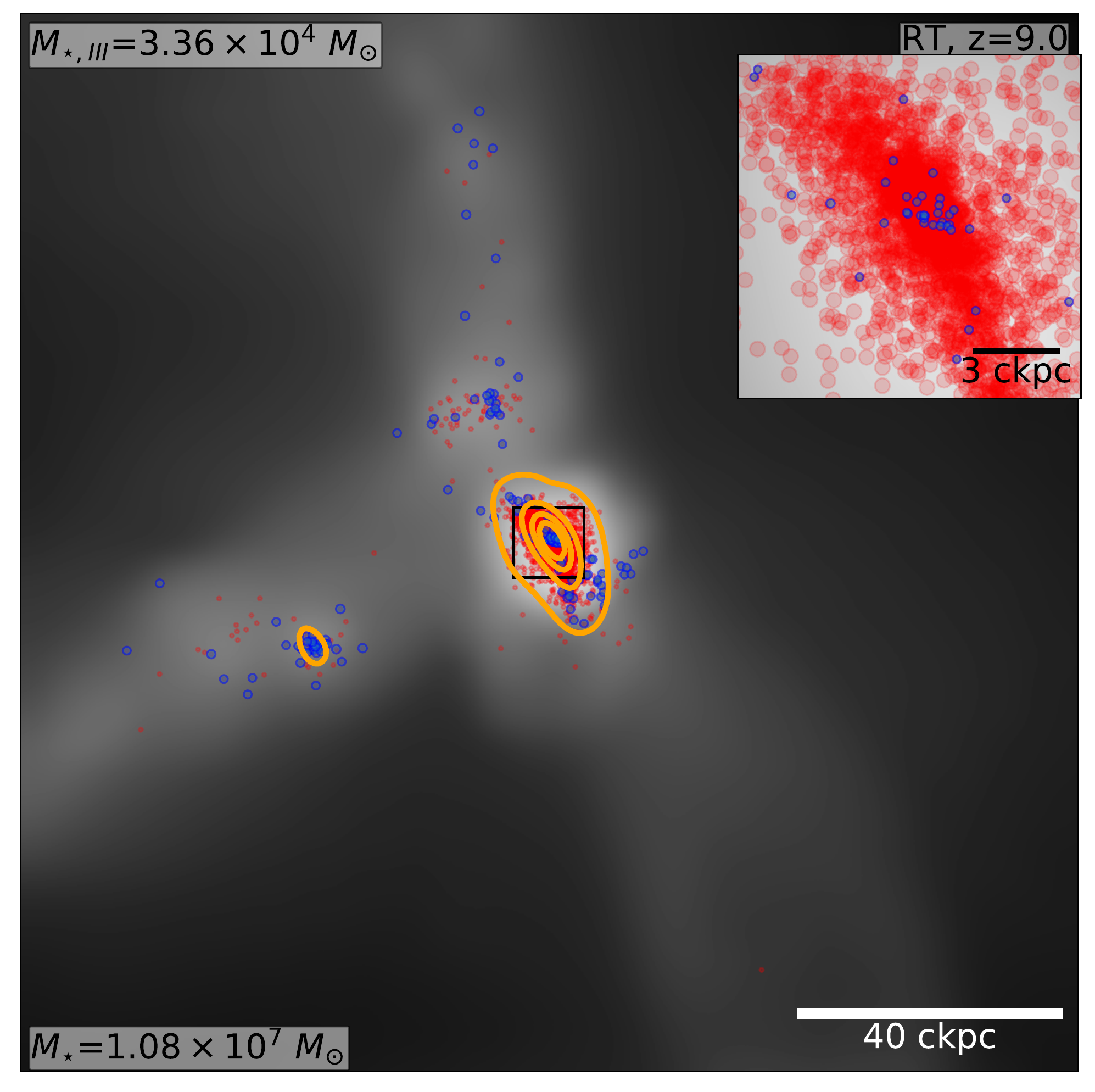}
\caption{Gas density and star particles (blue dots for Pop III stars, red for Pop II) along with orange contours indicating 20, 40, 60 \& 80\% of SP mass for a representative 160 ckpc region centered on a primordial halo at $12 \le z \le 9$ in each simulation. Insets depict the central 12 ckpc. Note that the \nonRTsim{} (cols 1 and 3) generates stars with significantly larger density contours than the \RTsim{} (cols 2 and 4). The \RTsim{} SP density profiles are largely concentrated in dense cores and along filaments as compared to the \nonRTsim{}, which displays a much more distributed stellar density structure. The gas density profile in the \RTsim{} is much smoother than that seen in the \nonRTsim{}. }
\label{fig:SFdenPlots}
\end{center}
\end{figure*}

Fig.~\ref{fig:popIIIyng} depicts $\xi'_2(r_k,t_l)$ for all the SPs in our simulations at $z=10$. 
Considering the spatial dimension ($y$-axis), the $\xi'_2(r_k,t_l)$ plot for the \nonRTsim{}, top left, displays SPs that smoothly approach a uniform distribution as $\Delta r \rightarrow 500$ ckpc, the top of the plot. Unsurprisingly, this captures the fact that star formation is concentrated in galaxy cores and on large scales the star particle distribution approaches uniformity.  Examining the time dimension ($x$-axis), on the other hand, we see the that star formation occurs in bursts of $\approx$20 Myr, followed by a gradual decline to $\Delta t  \rightarrow 50.$ This effect is most pronounced at $\Delta r \lesssim 1$ ckpc, indicating the typical scale of a star forming region is roughly this size. The plot allows us to see both the temporal and spatial suppression of star formation by SN in the case of the \nonRTsim{}.
 
The \RTsim{}/\nonRTsim{} plot, in the top right of this figure, depicts the \textit{relative} probabilities for SP spatial and temporal separations in the \RTsim{} as compared to the \nonRTsim{}.  In the central star forming regions at a scale $r < 1$ ckpc  (the blue region), we see relative suppression in the \RTsim{}. This simulation has a lower probability of forming stars as compared to the \nonRTsim{}.   This indicates that radiative heating and pressure has moved star formation in the \RTsim{} outside of the cores into the volume with $\Delta r \gtrsim 1$ ckpc.  This suppression can be a strong as a factor of $\approx 5$ over a timescale of $\gtrsim$ 20 Myrs. This timescale is in agreement with the typical extent of a star-forming burst (as identified in the upper left panel) in the \nonRTsim{}. It then becomes more moderate as $\Delta t \rightarrow 50,$ finally resulting in a mild enhanced probability of star-formation as the starburst fades away, the gas cools and become dense once again. 

Also in the \RTsim{}/\nonRTsim{} plot we see that the \RTsim{} generates SPs with a higher probability as compared to the \nonRTsim{} in the range $1\leq \Delta r \lesssim 10$ ckpc. This result, along with the relative suppression of SP formation in the halo cores, suggests that thermal pressure due to radiation is pushing star formation out to distances $\approx$1 ckpc.  The enhancement in this region correlates with the expected radius of the Str\" omgren ionization radius caused by stars in the \RTsim{}:
\begin{equation}
R_s = \left(\frac{3}{4 \pi} \frac{S_*}{n^2 \beta_2} \right) ^{1/3},
\label{eqn:strom}
\end{equation}
where $S_*$ is the stellar flux in ionizing photons per Solar mass, $n$ is the gas number density in units of H\, cm$^{-3}$, and $\beta_2 = 2\times 10^{-10}\, T_e^{-3/4}$ with units cm$^3$ s$^{-1}$ and is the recombination rate in terms of the effective temperature $T_e$. Assuming our stars generate $3\times 10^{47}$ ionizing photons per solar mass \citep{1999ApJS..123....3L,2003ApJ...589...35S,2011ascl.soft04003L}  this can be expressed as 
\begin{equation}
R_s = 2.3\, {\rm ckpc}\; \Phi^{1/3}\; n_1^{-2/3}\; T_4^{1/4}\; (1+z),
\label{eqn:strom2}
\end{equation}
where $\Phi$ is the star formation rate in units of solar masses per year, $n_1 = n/{\rm cm}^{-3}$, and $T_4 = \nicefrac{T_e}{10^4K}$, and the factor of (1+$z$) ensures we are in ckpc. For the largest halo in our \RTsim{} at $z=10$ we have $n_1=3$, $T_4=1$ and $\Phi$=0.022 computed between $z=12$ and $z=10,$ that results in $R_s=3.4$ ckpc, a little larger than the start of the relative increase in SP formation seen in Fig.~\ref{fig:popIIIyng}, as expected from values are taken from the most massive halo in the simulation. Hence we conclude that the thermal pressure from the ionized gas pushes most star formation out to approximately the Str\" omgren radius.

With these overall trends in mind, we turn to Pop III star formation.   Fig.~\ref{fig:popIIIyng}, lower left shows $\xi_2^{'}(r_k,t_l)$ for Classical Pop III stars in the \nonRTsim{}, which again are defined as those which form in simulation cells composed solely of unpolluted gas.  Here we see a nearly complete lack of Classical Pop III star formation after 10 Myr, the lifetime of massive stars, in volumes with $\Delta r \lesssim 1$ ckpc.  This polluted region in which Classical Pop III are suppressed then expands to $\approx$ 5 ckpc by 50 Myr, an average rate of 1 ckpc per 10 Myr or $\approx$ 100 km $s^{-1}$.  

Finally in Fig.~\ref{fig:popIIIyng}, lower right, we show $\xi_2^{'}(r_k,t_l)$ for the full Pop III star population including those that form in incompletely-mixed regions captured by our subgrid model. This figure displays continued Pop III star formation in the \nonRTsim{} that trails off smoothly as a function of the separation in both time and space.  Like the overall stellar population, Pop III stars are formed in star-forming regions with a typical scale of $\Delta r \lesssim 1$ ckpc, but  Pop III star formation trails off more rapidly in time, as metals mix and unpolluted gas becomes more scarce with time.

We can also compare the  $\xi_2^{'}(r_k,t_l)$ for Pop III stars to the ratio of $\xi_2^{'}(r_k,t_l)$ in the \RTsim{}/\nonRTsim{} to understand how suppression by radiative transfer will directly affect the Pop III stellar population.  The comparison shows that Pop III stars are not only clustered spatially on the scales that are suppressed by radiative feedback, but also clustered temporally on the time scale at which radiative feedback is most significant.  It is the combination of both types of clustering, then, that explains the strong differences in the Pop III star formation rate visible in Figs.~\ref{fig:sfrd} and \ref{fig:temps}.

We can also see the effect of radiative feedback on the location and morphology of this group of proto-galaxies in Fig.~\ref{fig:SFdenPlots}, which depicts the $z=12$ to $z=9$ evolution of the gas density and Pop III and Pop II stellar distributions in a 160 ckpc region around the densest section in the simulations.  At each redshift the \RTsim{} generates more compact proto-galaxies as exhibited in the extent of the enclosing density contours. By $z=10$ the \RTsim{}'s stellar mass is almost 2 times more compact than in the \nonRTsim{} and by $z=9$ the \RTsim{} galaxies are, for this example, at least 3 times more compact when considering the 80\% stellar mass contour. By this epoch, both simulations have generated almost the same mass in stars in this region. This supports the analysis of stellar clustering depicted using $\xi_2$ above. 

By $z=10$ the lower density gas is effectively removed from halos in the \RTsim{} leaving only the higher density filamentary structures as the areas that shield the gas from the effects of the radiation.  This initially results in enhanced star formation in regions typically at least 1 ckpc away from the initial cores. This is proto-galaxy satellite formation that is not seen in the \nonRTsim{}. Again, the \nonRTsim{} produces more connected structures. By $z=8$ we see that the \RTsim{} proto-galaxy has coalesced into a much more compact object that is approximately the same mass as its counterpart found in the \nonRTsim{}. However, 80\% of the mass of the \RTsim{} is contained in a region with a major axis $\approx 20$ ckpc while in the \nonRTsim{} the comparable mass is enclosed in a region almost 4 times that size.

\section{Conclusions} 

The next decade will bring unparalleled advances in the observational study of the first stars and galaxies, but realizing the full potential of these observations requires simulations that capture the key physical processes that affected these objects.  Here we have used cosmological simulations to demonstrate the importance of modeling both radiative transfer and turbulent mixing when studying primordial galaxy morphology, composition, and growth. 

By comparing simulations with and without a detailed model for radiative transfer but with exactly the same physical parameters and turbulent mixing model, we are able to directly quantify how radiative transfer and mixing work together to determine the evolution of Pop III stars and galaxies.  Of course, radiative transfer has an impact on all stars, and the star formation rate density in the \RTsim{} is lower at high-redshift and higher at low-redshift, as compared to the \nonRTsim{}  that assumes that reionization occurs instantaneously at $z=9.$  This difference is due to the additional thermal pressure that ionized pockets of gas feel at high redshift in the  \RTsim{}, which is not captured in the \nonRTsim{}.

But the most interesting impact of radiative transfer is on the evolution of Population III stars.  In fact, the overall PopIII star formation rate density is suppressed by a factor of $\approx 6$ in the  \RTsim{} at all redshifts, both before and after $z=9,$ where the \nonRTsim{} assumes instantaneous reionization.  This indicates that understanding the simultaneous propagation of ionization fronts and mixing of heavy elements is essential to understanding Pop III evolution, and that radiative transfer is necessary to capture the detailed properties of this interplay. 

Likewise, an assumption that all polluted regions are instantaneously mixed leads to an inaccurate picture: identifying less than 10\% of total number of Pop III stars in the simulations (at the resolution used), and failing to capture the suppression of Pop III star formation that occurs as radiative feedback slows gas accretion and provides more time for mixing to pollute a larger fraction of star-forming gas.   Note that even high-resolution simulations like ours cannot capture this suppression in Pop III stars in the \RTsim{} without a subgrid model of the mixing time required to distribute metals throughout a given region. It is simply not possible with the current generation of super computers to  simulate a representative cosmological volume through the full process of reionization while still tracking the formation of individual stars.

To better understand the impact of radiative transfer on Pop III star-formation, we have made use of the two-point temporal-spatial correlation function $\xi'_2(r_k,t_l),$ which quantifies the excess probability of forming stars with a given separation in time and space. This measure confirms and quantifies that star formation in proto-galaxies proceeds more rapidly in the halo cores of the \nonRTsim{} as compared to the \RTsim{}, and that supernova feedback alone does not quench star formation as effectively as the combination of SN and radiative feedback in regions of size $\lesssim 1$ ckpc and at times $\lesssim $25 Myr. Hence the cores of young galaxies in the \RTsim{} experience a relative suppression of star formation in the first $\approx$ 25 Myr after a starburst during which radiative feedback from massive stars is strongest, followed by an enhanced probability of forming stars at $\approx$ 40 Myr as massive stars die and the UV flux drops dramatically allowing gas to reaccrete.  

Note that the $\approx 1$ ckpc size of the region in which this suppression and reaccrection occur correlates with the Str\" omgen radius for the typical density and stellar mass of a high-redshift starbursting proto-galaxy.  At distances larger than the Str\" omgen radius, on the other hand, we find that radiative feedback pushes gas out to distances between 1 and 10 ckpc, encountering the dense filaments feeding the galaxy. It is here that we find on-going star formation in the \RTsim{} that is enhanced to a rate 4-10x higher than in the same region in the \nonRTsim{}. This results in numerous, low mass satellites at $z \ge 10,$ which are spatially separated to a far higher degree than stars making up the \nonRTsim{} galaxies. 

Relating these changes to metal mixing helps explain the impact of radiative transfer on Pop III evolution.  If we limit ourselves to the ``Classical" Pop III,  which form in simulation zones in which there are no metals, then the impact of radiative transfer is minimal.  These are the very first stars to form in a new burst of star formation, and they are quickly extinguished as soon as SNe begin to enrich the medium.  Accounting for the time it takes metals to mix into the gas, on the other hand, completely changes the picture. Pop III stars formed in areas of active mixing are formed continually out to 50 Myr after the start of a starburst, and hence they are strongly impacted by radiative feedback. 

Lastly, we examine impact of radiative transfer on galaxy morphology. Without RT and the associated heating/ionization of the gas out to the Str\" omgren radius, the \nonRTsim{} generates far more stars in the central ckpc of halos than  the \RTsim{}. At the same time, the  \nonRTsim{}  generates a much smoother and extended stellar density profile, while more isolated low mass satellites are seen in the \RTsim{}. By $z=8$ the initially unused, heated gas in the \RTsim{} has mostly cooled and is converted into stars, resulting in galaxies, counting the mass of the satellites, of comparable mass in both simulations. Given the short lifetimes of massive Pop III stars, the final composition of the galaxies in the two simulations converges at these later epochs. However, the lasting impact of these two growth paths is the morphology of the galaxies, and the \nonRTsim{} galaxies remain more extended at all redshifts we studied. 

With the recent launch of {\em JWST}, we are now on the cusp of a momentous change in our observational understanding of the first stars and galaxies.  Here we have shown that fully capitalizing on these advances will require a new generation of cosmological simulations that include both radiative transfer and turbulent mixing. Without radiative transfer simulations will over-estimate the prevalence of Pop III stars, and without mixing  simulations will vastly underestimate the Pop III content of galaxy cores. The net effect of modeling radiative transfer and subgrid mixing is a fourfold increase in the mass-fraction of Pop III stars at $9 \le z \le 13$ as compared to simulations that do not include these physical processes. Further, modeling RT is important for understanding the morphology of early galaxies. Stellar radiation from the first stars results in galaxies that are more compact and less luminous than those generated by non-RT simulations. Given our results, we predict that future JWST observations will bear-out these qualitative properties. These properties, in turn, will have a direct impact on predictions of the luminosities and surface brightness profiles of the galaxies most likely to contain Pop III stars. We will explore these observational properties in a future publication using a significantly larger simulation volume.

\acknowledgments
We would like to thank Joakim Rosdahl at the Centre de Recherche Astrophysique de Lyon for very useful discussions concerning \textsc{Ramses-RT}. We also thank Paul Shapiro at the University of Texas, Austin for helpful discussions about the first stars and early galaxies and their role in reionization. This work was supported by the United States Naval Academy (USNA) and NSF Grant AST-1715876. The simulations for this work were carried out on the USNA Advanced Research Cluster (ARC) and on the Pittsburgh Supercomputing Center (PSC) Bridges2 Supercomputer using award PHY200095. 

\software{\textsc{ramses} \cite{2010ascl.soft11007T}, MUSIC \citep{2013ascl.soft11011H}, pynbody \cite{2013ascl.soft05002P}}

\clearpage


%
\bibliographystyle{apj}
\bibliography{PopIII_RT_v_nonRT.bib}
\end{document}